\documentclass[reprint,amsmath,amssymb,aps,pra,showkeys,showpacs]{revtex4-1}
\usepackage{graphicx}

\newcommand{\ket}[1]{\mathinner{|{#1}\rangle}}
\newcommand{\bra}[1]{\mathinner{\langle{#1}|}}

\begin{document}
\title{Lindblad Formalism based on Fermion-to-Qubit mapping for Non-equilibrium Open-Quantum Systems}
\author{Fabr\'{i}cio M. Souza}
\email{fmsouza@ufu.br}
\author{L. Sanz}
\affiliation{Instituto de F\'isica, Universidade Federal de Uberl\^andia, 38400-902 Uberl\^andia, MG, Brazil}
\date{\today}

\begin{abstract}
We present an alternative form of master equation, applicable on the analysis of non-equilibrium dynamics of fermionic open quantum systems. The formalism considers a general scenario, composed by a multipartite quantum system in contact with several reservoirs, each one with a specific chemical potential and in thermal equilibrium. With the help of Jordan-Wigner transformation, we perform a fermion-to-qubit mapping to derive a set of Lindblad superoperators that can be straightforwardly used on a wide range of physical setups.To illustrate our approach, we explore the effect of a charge sensor, acting as a probe, over the dynamics of electrons on coupled quantum molecules. The probe consists on a quantum dot attached to source and drain leads, that allows a current flow. The dynamics of populations, entanglement degree and purity show how the probe is behind the sudden deaths and rebirths of entanglement, at short times. Then, the evolution leads the system to an asymptotic state being a statistical mixture. Those are signatures that the probe induces dephasing, a process that destroys the coherence of the quantum system.
\end{abstract}
\maketitle

\section{Introduction}
\label{sec:intro}
The study of the boundary between quantum and classical mechanics raised as one of the most interesting
and challenging issues for the last thirty years~\cite{Zurek91,*HarochePT98,Schlosshauerbook,Schlosshauer05,*Schlosshauer2014}.
Motivated by the necessity of exploring problems like the measurement of quantum properties or the classical limit for a specific quantum model,
different ways to treat the so-called \textit{open quantum systems} has been proposed~\cite{Schlosshauerbook,Breuerbook}.
In particular, the density matrix formalism~\cite{Fano57} becomes an important theoretical frame to explore multipartite
systems with mixed quantum and classical features. From the wide set of problems linked with open systems,
the analysis of the dynamics of a quantum system in contact with \textit{reservoirs}, larger physical systems,
stands as a fundamental quest. Speaking specifically of the ``know-how", master equations become adequate to circumvent the task,
being deduced by tracing out the variables of the reservoirs~\cite{Schlosshauerbook,Breuerbook}. The integro-differential
equations obtained after the application of several approximations permits to summarize the effect of reservoirs via a
Lindblad operator~\cite{Gorini76,Lindblad76}. Extensions has been made in order to include memory effects,
known as non-markovian approaches~\cite{zhang2012general,BreuerRMP16,*Scorpo17,*Bernardes2017},
which has been observed in carefully prepared experimental setups~\cite{Li11,*liu2011experimental}.

On the other hand, since the seminal work of Jauho, Wingreen and Meir~\cite{Jauho94} about non-equilibrium quantum transport,
a wealth of theoretical and experimental works have investigated the transport phenomena under the action of time-vary fields~\cite{platero2004photon,*cota2005ac,souza2007spin,*souza2007transient,trocha2010beating,*perfetto2010correlation,Assuncao13,*odashima2017time}.
Recently, open quantum systems out-of-equilibrium have been theoretically investigated in quantum dots attached to leads in the presence of
photonic or phononic fields~\cite{liu2014photon,*kulkarni2014cavity,*hartle2015effect,*purkayastha2016out,*agarwalla2016tunable,*reichert2016dynamics,*mann2016dissipative}.
In this specific context, a method to deal with transport problems in semiconductor nanoestrutures has been developed by W.-M. Zhang
and co-workers~\cite{Yang17,*Xiong15,Zhang12prl,Jin10}, mixing the density operator formalism with nonequilibrium Green functions. From the point of view of the treatment of time-memory effects, this method is powerful because its direct application on the description of non-markovian setups. Still, the approach requires some familiarity with Keldysh non-equilibrium Green function technique.

Here, we present a formalism that offers an alternative path with immediate application on the study of dynamics of a general configuration of open quantum systems far from equilibrium. Using the Fermion-to-Qubit (FTQ) mapping, we provide a straightforward recipe to construct both, multi-partite Hamiltonian and Lindbladians, as tensor products of Pauli matrices. The FTQ mapping also sets automatically the complete computational basis to analyze the dynamics, written in terms of occupied and non-occupied states. The formalism presented here opens the possibility of future applications in the context of fermionic quantum computation~\cite{Bravyi02}.

The paper is organized as follows: in Sec.~\ref{sec:genform}, we set the foundations of our formalism, starting with the definition of a general form for fermionic operators. It is deduced an expression for a generic reservoir-system coupling, which is the key behind the construction of super-operators for open quantum dynamics. Section~\ref{sec:noninteract} presents the deduction of Lindbladian super-operators, for the case of non-interacting reservoirs considering a markovian condition. Section~\ref{sec:application} is devoted to the discussion of an application of our formalism on the context of transport phenomena. We focus on the behavior of electrons on charged quantum molecules, being probed by a nearby narrow conduction channel describing the action of a charge sensor. The behavior of populations, the entanglement dynamics, and the purity permits to conclude that the probe induces dephasing, a decoherence process which acts over the quantum dynamics of the coupled molecules. In Sec.~\ref{sec:summary} we summarize our results.

\section{Fermion-to-qubit mapping and the general Hamiltonian}
\label{sec:genform}
\begin{figure}[tb]
\centering\includegraphics[width=1\linewidth]{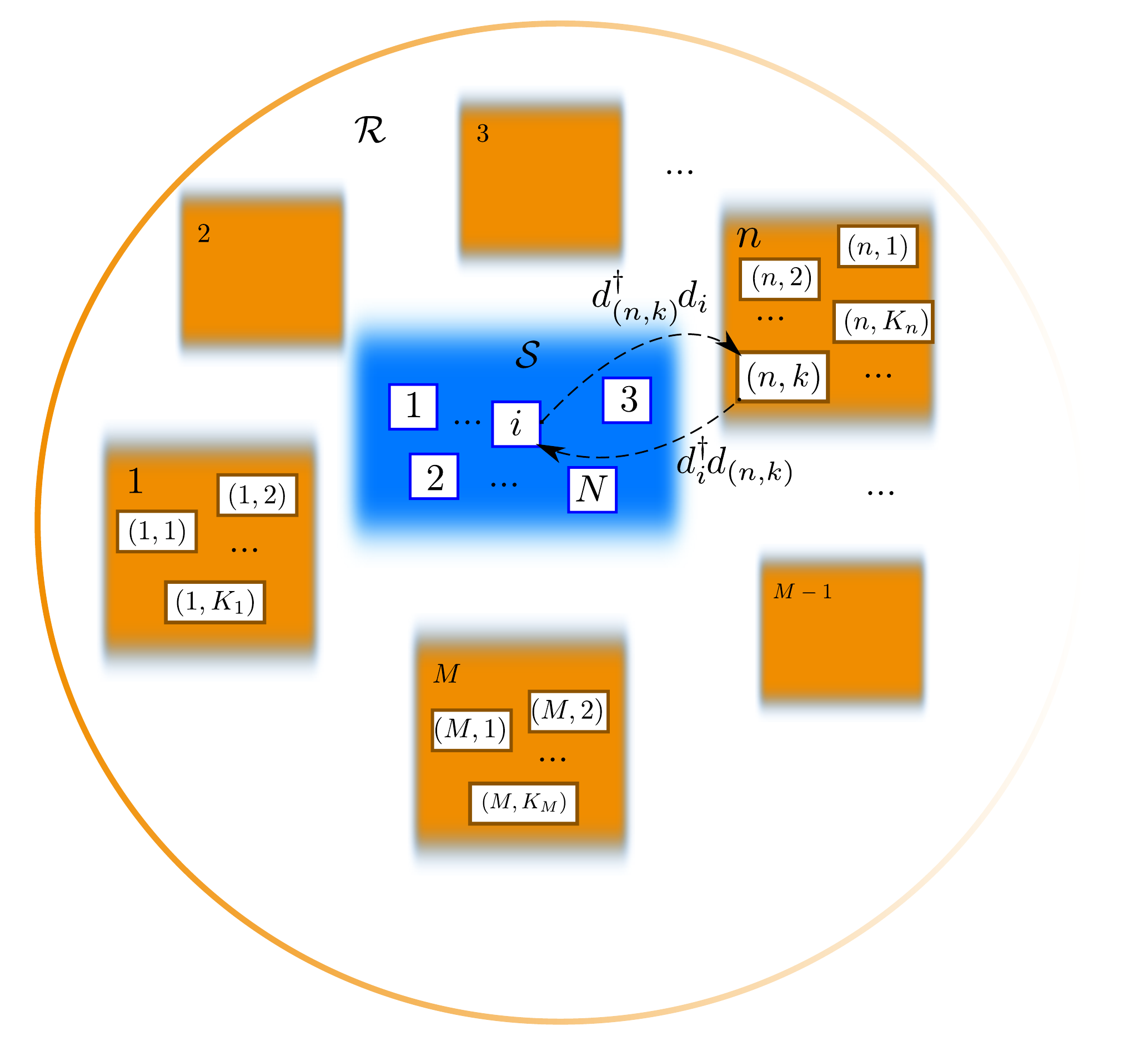}
\caption{Illustration of a generic setup composed of a multipartite system $\mathcal{S}$ interacting with $M$ reservoirs.}
\label{fig:system}
\end{figure}

Consider the open multipartite system illustrated in Fig.~\ref{fig:system} with $N$ subsystems in space $\mathcal{S}$, in contact with $M$ reservoirs, each with $K_n$ inner states, defined in a space denoted as $\mathcal{R}$. We use $i$ as the index of the $i$-th subsystem in $\mathcal{S}$ so $i=1,2..,N$. In the reservoir space, we use two indexes: $n$, which labels the $n$-th reservoir, with $n=1,...,M$, and $k$, indicating the $k$-th state with $k=1,...,K_n$. The dimension of the whole, system and reservoirs, is given by $D=N+\sum_{n=1}^{M} K_n$.
The general Hamiltonian can be written as $H(t)=H_0+V(t)$ where
\begin{eqnarray}
\label{eq:Hamilterms}
H_0&=&H_\mathcal{S} \otimes I_\mathcal{R}^{\otimes \left(D-N\right)} + I_\mathcal{S}^{\otimes N} \otimes H_{\mathcal{R}},\nonumber\\
V(t)&=&\sum_{i=1}^{N}\sum_{n=1}^{M}\sum_{k=1}^{K_n}u_n(t)V_{i,(n,k)}d^{\dagger}_{(n,k)}d_i+\mathrm{h. c.}
\end{eqnarray}
where $H_{\mathcal{S}(\mathcal{R})}$ is the Hamiltonian of the multipartite system (reservoirs)
without coupling and the term $V(t)$ describes the coupling as a hopping process:
a particle is annihilated ($d_i$) at $\mathcal{S}$ at the same time that it is created ($d_{(n,k)}^\dagger$)
in $\mathcal{R}$ and vice versa. Inside the coupling term, the function $u_n(t)$ is a time-dependent parameter~\cite{Jauhobook}, and  $V_{i,(n,k)}$ provides the coupling strength between system and reservoirs.

Now we apply the FTQ mapping to this general system.
The operator $d_m$ is defined by using the Jordan-Wigner transformation~\cite{Schallerbook} as~\footnote{This tool has been applied in problems involving interacting quantum dynamics, one being the classic work of Haldane~\cite{Haldane80}
who show the equivalence between the $XXZ$ model for a Heisenberg-Ising chain and interacting spinless fermions.
Other example of the application of this transformation on the physics of strong correlated systems are the use
of a generalized of the transformation in the solution of some cases of lattice models~\cite{Batista01}.}
\begin{equation}
d_m=\sigma_z^{\otimes\left(m-1\right)}\otimes\sigma_{-}^{(m)}\otimes I^{\otimes\left(D-m\right)},
\label{eq:mapdm}
\end{equation}
where the index $m$ now runs over both, $\mathcal{S}$ and $\mathcal{R}$ indexes, $\sigma_z$ is a Pauli matrix, $I$ is the $2\times2$ identity matrix, and $O^{\otimes L}$ indicates a succession of $L$ tensorial products of operator $O$. The creation operator $d^{\dagger}_m$ is obtained by replacing $\sigma_-^{(m)}$ by $\sigma_+^{(m)}$ with $\sigma_{\pm}=(\sigma_x \pm i \sigma_y)/2$. It is straightforward to prove that $d_m$ and $d_m^\dagger$ follows the anticommutation relations $\left\{d_m,d^{\dagger}_l\right\}=\delta_{m,l}$, $\left\{d_m,d_l\right\}=\left\{d^{\dagger}_m,d^{\dagger}_l\right\}=0$.

The explicit form for $V(t)$ is now written as:
\begin{eqnarray}
\label{eq:V}
V(t)&=&\sum_{i,n,k} u_n (t)V_{i,(n,k)}\left(S_i \otimes \sigma_z^{\otimes K_1} \otimes \sigma_z^{\otimes K_2} \otimes \cdot \cdot \cdot \otimes \right. \\
 &&\left. \sigma_z^{\otimes K_{n-1}} \otimes  R_{n,k}^\dagger \otimes I^{\otimes K_{n+1}} \cdot \cdot \cdot \otimes I^{\otimes K_M}\right) +\mathrm{h. c.},\nonumber
\end{eqnarray}
where
\begin{eqnarray}
\label{eq:SRoperators}
S_i&=&I^{\otimes\left(i-1\right)}\otimes\sigma_-^{(i)}\otimes\sigma_z^{\otimes\left(N-i\right)},\\
R_{n,k}^{\dagger}&=&\sigma_z^{\otimes k-1} \otimes \sigma_+^{(k)} \otimes I^{\otimes K_n - k}\nonumber.
\end{eqnarray}
The $S_i$ ($R_{n,k}^{\dagger}$) operators run only over the subspace $\mathcal{S}$ ($n$-th subspace of $\mathcal{R}$), and they preserve the fermionic anticommutation relations.

In order to describe the quantum evolution of the entire $D$ dimensional system, we write the Von Neumann equation
in the interaction picture, $\dot{\rho}(t)= \mathcal{L}(t) \rho_0 + \int_{0}^t dt_1 \mathcal{L}(t) \mathcal{L}(t_1)\rho(t_1)$
where $\mathcal{L}(t)$ is the Liouvillian superoperator,
$\mathcal{L}(t)\rho(t_1)=-i [V_I(t),\rho(t_1)]$ ($\hbar=1$), and $V_I(t)$
is the coupling term defined as $V_I(t)=e^{iH_0t}V(t)e^{-iH_0t}$
with $e^{iH_0t}=e^{iH_\mathcal{S}t}\otimes e^{iH_\mathcal{R}t}$.
This transformation applies over the tensorial product inside
Eq.~(\ref{eq:V}) resulting on a tensorial product between system operators $S_i(t)=e^{iH_\mathcal{S}t}S_ie^{-iH_\mathcal{S}t}$
and similar terms for the reservoirs. Notice that, at the moment,
we are treating the full form of the system-reservoir interaction,
using a mathematical tool to distinguish the system from the reservoirs without a loss of generality.

\section{Non-interacting reservoirs and the Markov approximation}
\label{sec:noninteract}
Let us assume the Born approximation, $\rho(t)=\rho_\mathcal{S}(t) \otimes \rho_{\mathcal{R}}$,
where $\rho_\mathcal{S}(t)$ is the reduced density matrix
of multipartite system and $\rho_{\mathcal{R}}$ are the
density matrices for reservoirs, thus
$\rho_{\mathcal{R}}=\rho_1 \otimes \rho_2 \otimes \cdot \cdot \cdot \otimes \rho_M$.
We now consider the effect of non-interacting reservoirs, set in thermodynamical equilibrium, each described by
Hamiltonian $H_{\mathcal{R}}=\bigoplus_{n=1}^M H^{(n)}$ with $H^{(n)}=\bigoplus_{k=1}^{K_n}H^{(n,k)}=\bigoplus_{k=1}^{K_n} \varepsilon_n^{(k)} (\sigma_+ \sigma_-)^{(k)}$ and $\varepsilon_n^{(k)}$ being the energy of the $k$-th mode of reservoir $n$.
The density matrix for each reservoir is a mixed state described by
$\rho_n = \frac{1}{Z} \mathrm{Exp}[-\beta \bigoplus_{k=1}^{K_n} \epsilon_{n}^{(k)} (\sigma_+ \sigma_{-})^{(k)}]$,
where $\beta=1/(k_B T)$, $\epsilon_{n}^{(k)}$ is the free particle energy measured
from the chemical potential $\mu_n$, and $Z$ is the partition function.
After taking the partial trace over reservoirs degrees of freedom and
ignoring the null terms, we find
\begin{widetext}
\begin{eqnarray}
\label{eq:fulldotrho}
 \dot{\rho}_\mathcal{S}(t)&=&-\frac{1}{2\pi} \int_0^t dt_1 \sum_{i,j,n,k} \large{\{}\Gamma_{i,j}^{n,k}(t,t_1)S_i(t)S_j^\dagger(t_1) \rho_\mathcal{S}(t_1)
  \mathrm{Tr}_{\mathcal{R}}[ \rho_1 \otimes \rho_2 \otimes \cdot \cdot \cdot \otimes \rho_{n-1} \otimes
 R_{n,k}^\dagger(t) R_{n,k}(t_1) \rho_n \otimes \cdot \cdot \cdot \otimes \rho_{M}  ]\\
& +&\Gamma_{i,j}^{n,k}(t_1,t)S_i^\dagger(t) S_j(t_1) \rho_\mathcal{S}(t_1)
 \mathrm{Tr}_{\mathcal{R}}[ \rho_1 \otimes \rho_2 \otimes \cdot \cdot \cdot \otimes \rho_{n-1} \otimes
 R_{n,k}(t) R_{n,k}^\dagger(t_1) \rho_n \otimes \cdot \cdot \cdot \otimes \rho_{M}]\nonumber \\
&-&\Gamma_{i,j}^{n,k}(t_1,t) S_j(t_1) \rho_\mathcal{S}(t_1) S_i^\dagger(t)
 \mathrm{Tr}_{\mathcal{R}}[\sigma_z^{\otimes K_1} \rho_1 \sigma_z^{\otimes K_1} \otimes\cdot \cdot \cdot \otimes \sigma_z^{\otimes K_{n-1}} \rho_{n-1} \sigma_z^{\otimes K_{n-1}}\otimes R_{n,k}^\dagger(t_1) \rho_n R_{n,k}(t)  \otimes \rho_{n+1}\cdot \cdot \cdot \otimes \rho_{M}]\nonumber \\
&-&\Gamma_{i,j}^{n,k}(t,t_1)S_j^\dagger (t_1) \rho_\mathcal{S}(t_1) S_i(t)
 \mathrm{Tr}_{\mathcal{R}}[ \sigma_z^{\otimes K_1} \rho_1 \sigma_z^{\otimes K_1} \otimes \cdot \cdot \cdot \otimes \sigma_z^{\otimes K_{n-1}} \rho_{n-1} \sigma_z^{\otimes K_{n-1}}\otimes R_{n,k}(t_1) \rho_n R_{n,k}^\dagger(t)  \otimes \rho_{n+1} \cdot \cdot \cdot \otimes \rho_{M}] \nonumber\\
&+& h.c. \large{\}},\nonumber
\end{eqnarray}
\end{widetext}
where $R^{(\dagger)}_{n,k}(t)=e^{iH^{(n,k)}t}R^{(\dagger)}_{n,k}e^{-iH^{(n,k)}t}$ and $ \Gamma_{i,j}^{n,k}(t,t_1) = 2 \pi V_{i,(n,k)} V_{j,(n,k)}^\star u_n(t) u_n^\star(t_1)$. The equation for the reduced density matrix of the system is written as
\begin{eqnarray}
\label{eq:rhodotbftint}
\dot{{\rho}}_\mathcal{S}(t)&=&-\frac{1}{2\pi} \sum_{i,j}\sum_{n,k}\Big\{f_{n,k} \int_0^t dt_1\Gamma_{i,j}^{n,k}(t,t_1)e^{i \varepsilon_n^{(k)} (t-t_1)} \\
&&\times\left[{S}_i(t){S}_j^\dagger(t_1){\rho}_\mathcal{S}(t_1)-{S}_j^\dagger (t_1) {\rho}_\mathcal{S}(t_1) {S}_i(t)\right]\nonumber \\
&&+(1-f_{n,k})\int_0^t dt_1\Gamma_{i,j}^{n,k}(t_1,t)  e^{-i \varepsilon_n^{(k)} (t-t_1)}\nonumber\\
&&\times \left[{S}_i^\dagger(t) {S}_j(t_1) {\rho}_\mathcal{S}(t_1)-{S}_j(t_1) {\rho}_\mathcal{S}(t_1) {S}_i^\dagger(t)\right]+ h.c. \Big\},\nonumber
\end{eqnarray}
after using the Baker-Hausdorff Lemma on the calculation of  $\mathrm{Tr}_n \{ R_{n,k} (t_1) \rho_n R_{n,k}^\dagger (t) \} = f_{n,k} e^{i \varepsilon_n^{(k)} (t-t_1)}$ and $\mathrm{Tr}_n \{ R_{n,k}^\dagger (t_1) \rho_n R_{n,k} (t) \} = (1-f_{n,k})e^{-i\varepsilon_n^{(k)}(t-t_1)}$, where $f_{n,k}=1/[1+e^{\beta \epsilon_n^{(k)}}]$
is the Fermi distribution function for reservoir
$n$~\footnote{Additionally, $\mathrm{Tr}_n\{ \rho_n  \}=\mathrm{Tr}_n\{ \sigma_z^{\otimes K_n} \rho_n \sigma_z^{\otimes K_n } \}=1$}.
At this point, we consider the wide-band limit~\cite{Jauhobook}.
We set $V_{i,(n,k)}=V_{i,n}$, meaning that all inner states on reservoir $n$ have the same coupling with the system.
The sum over $k$ turns into an integral, $\sum_k[...]\rightarrow \int[...] \mathfrak{D}_n(\varepsilon)d\varepsilon$,
where the density of states is assumed constant, $\mathfrak{D}_n(\varepsilon)=\mathfrak{D}_n$.
The bias voltage $eV$ is given by $\mu_n-\mu_m =eV$, where the source (drain) chemical potential
is $\mu_n$ ($\mu_m$). If $eV$ is high and considering low values of temperature,
we can assume that $f_{n(m),k}=f_{n(m)}=1(0)$ resulting in,
\begin{equation}
\label{rhoint2}
\dot{{\rho}}_{S}(t)=-\frac{1}{2} \sum_{i,j,n}  \Gamma_{i,j}^{(n)}(t) \left[F_n(t,i,j)+G_n(t,i,j)\right],
\end{equation}
with
\begin{eqnarray}
\label{rhoint2p2}
F_n(t,i,j)&=& f_n\left[ {S}_i (t){S}_j^\dagger (t){\rho}_\mathcal{S} (t)-{S}_i^\dagger (t) {\rho}_\mathcal{S} (t) {S}_j (t)\right.\\
&&\left.+{\rho}_\mathcal{S} (t) {S}_j (t) {S}_i^\dagger (t) - {S}_j^\dagger (t) {\rho}_\mathcal{S} (t) {S}_i (t)\right]\nonumber \\
G_n(t,i,j)&=&\left(1-f_n\right)\left[{S}_i^\dagger (t) {S}_j (t) {\rho}_\mathcal{S} (t) - {S}_i (t)  {\rho}_\mathcal{S} (t) {S}_j^\dagger (t)\right.\nonumber\\
&&\left. +{\rho}_\mathcal{S} (t) {S}_j^{\dagger} (t) {S}_i(t) - {S}_j(t) {\rho}_\mathcal{S} (t) {S}_i^{\dagger} (t)\right],\nonumber
\end{eqnarray}
and $\Gamma_{i,j}^{(n)}(t)=2\pi V_{i,n}V^{*}_{j,n}\mathfrak{D}_n\left|u_n(t)\right|^2$.
We apply the $\mathbf{vec}$ operation~\cite{Havel03} to both sides of Eq.~(\ref{rhoint2}),
obtaining $\frac{d}{dt} \mathbf{vec}[{\rho}_\mathcal{S}] (t) = - \frac{1}{2}  {\mathbf{\Gamma}}(t)  \mathbf{vec}[ {\rho}_\mathcal{S}] (t)$,
where ${\mathbf{\Gamma}}(t)$ is $N^2\times N^2$ supermatrix defined as
\begin{eqnarray}
\label{eq:supmatgamma}
{\mathbf{\Gamma}}(t) &=& \sum_{i,j,n}  \Gamma^{(n)}_{i,j}(t) \left( f_n\left\{I_\mathcal{S} \otimes{S}_i (t) {S}_j^\dagger (t)- {S}^T_j (t) \otimes  {S}_i^\dagger (t) \right.\right.\nonumber \\
&&\left.\left. +[{S}_j (t){S}_i^\dagger(t)]^T \otimes I_\mathcal{S} - {S}^T_i(t)\otimes{S}_j^\dagger(t)\right\}\right.\nonumber\\
&&\left.+(1-f_n)\left\{ I_\mathcal{S} \otimes {S}_i^\dagger (t){S}_j (t) - {S}_j^{\dagger T} (t)\otimes{S}_i (t)\right.\right.\nonumber \\
&&\left.\left. +[{S}_j^\dagger(t){S}_i (t)]^T \otimes I_\mathcal{S} - {S}_i^{\dagger T}(t)\otimes{S}_j (t) \right\}\right),
\end{eqnarray}
where the superscript $T$ means matrix transposition.

Writing the reduced density matrix in the Schr\"odinger picture,
$\mathbf{vec}[\rho^s_\mathcal{S}] (t) = [\mathrm{Exp}(i H_\mathcal{S}^T t) \otimes \mathrm{Exp}(-i H_\mathcal{S} t) ] \mathbf{vec}[ {\rho}_\mathcal{S}] (t)$,
and taking the time derivative we find the differential equation,
\begin{equation}\label{eq:drhodt}
 \frac{d}{dt}\mathbf{vec}[\rho^s_\mathcal{S}] (t) = \mathcal{L} (t) \mathbf{vec}[\rho^s_\mathcal{S}] (t),
\end{equation}
where the superoperator $\mathcal{L}$ is given by
\begin{equation}
\label{eq:rhodotshro}
\mathcal{L} (t)= \mathcal{L}_0 - \frac{1}{2} \sum_{i,j,n}\Gamma^{(n)}_{i,j}(t)\left[f_n\mathcal{L}^{+}_{ij}+\left(1-f_n\right)\mathcal{L}^{-}_{ij}\right],
\end{equation}
with
\begin{eqnarray}
\label{eq:rhodotshrop2}
 \mathcal{L}_0 &=& -i (I_\mathcal{S} \otimes H_\mathcal{S} - H_\mathcal{S}^T \otimes I_\mathcal{S})\nonumber\\
 \mathcal{L}^{+}_{ij}&=&I_\mathcal{S} \otimes S_iS^{\dagger}_j+S_iS^{\dagger}_j\otimes I_\mathcal{S}-S^{\dagger}_i\otimes S^{\dagger}_j-S^{\dagger}_j\otimes S^{\dagger}_i\nonumber\\
 \mathcal{L}^{-}_{ij}&=&I_\mathcal{S} \otimes S^{\dagger}_i S_j+S_i^{\dagger}S_j\otimes I_\mathcal{S}-S_i\otimes S_j-S_j\otimes S_i.\nonumber
\end{eqnarray}
Eq. (\ref{eq:drhodt}) has the formal solution
\begin{equation}\label{rhointegrated}
 \mathbf{vec}[\rho^s_S] (t) = \mathcal{T} e^{\int_0^t d\tau \mathcal{L}(\tau)} \mathbf{vec}[\rho^s_S] (0),
\end{equation}
where $\mathcal{T}$ is the chronological time-ordering operator.
Writing the superoperators above in terms of Pauli matrices, we arrive to
\begin{widetext}
\begin{eqnarray}
\label{eq:Lijsigma}
\mathcal{L}^{\pm}_{i=j}&=&I^{\otimes (N+i-1)} \otimes (\sigma_{\mp} \sigma_{\pm})^{(i)} \otimes I^{\otimes (N-i)}+ I^{\otimes (i-1)} \otimes (\sigma_{\mp} \sigma_{\pm})^{(i)} \otimes I^{\otimes (2N-i)}\\
&&-2 I^{\otimes (i-1)} \otimes \sigma_{\pm}^{(i)} \otimes \sigma_z^{\otimes (N-i)}\otimes I^{\otimes (i-1)} \otimes \sigma_{\pm}^{(i)} \otimes \sigma_z^{\otimes (N-i)},\nonumber\\
\mathcal{L}^{\pm}_{i<j}&=&\pm I^{\otimes (N+i-1)} \otimes \sigma_{\mp}^{(i)}\otimes\sigma_z^{\otimes (j-i-1)}\otimes \sigma_{\pm}^{(j)} \otimes I^{\otimes (N-j)}
\pm I^{\otimes (i-1)} \otimes \sigma_{\mp}^{(i)}\otimes\sigma_z^{\otimes (j-i-1)}\otimes \sigma_{\pm}^{(j)} \otimes I^{\otimes (2N-j)}\nonumber\\
&-&I^{\otimes (i-1)} \otimes \sigma_{\pm}^{(i)} \otimes \sigma_z^{\otimes (N-i)} \otimes I^{\otimes (j-1)} \otimes \sigma_{\pm}^{(j)} \otimes \sigma_z^{\otimes (N-j)}
-I^{\otimes (j-1)} \otimes \sigma_{\pm}^{(j)} \otimes \sigma_z^{\otimes (N-j)} \otimes I^{\otimes (i-1)} \otimes \sigma_{\pm}^{(i)} \otimes \sigma_z^{\otimes (N-i)},\nonumber\\
\mathcal{L}^{\pm}_{i>j}&=&\pm I^{\otimes (N+j-1)} \otimes \sigma_{\pm}^{(j)}\otimes\sigma_z^{\otimes (i-j-1)}\otimes \sigma_{\mp}^{(i)} \otimes I^{\otimes (N-i)}\pm I^{\otimes (j-1)} \otimes \sigma_{\pm}^{(j)}\otimes\sigma_z^{\otimes (i-j-1)}\otimes \sigma_{\mp}^{(i)} \otimes I^{\otimes (2N-i)}\nonumber \\
&-&I^{\otimes (i-1)} \otimes \sigma_{\pm}^{(i)} \otimes \sigma_z^{\otimes (N-i)} \otimes I^{\otimes (j-1)} \otimes \sigma_{\pm}^{(j)} \otimes \sigma_z^{\otimes (N-j)}-I^{\otimes (j-1)} \otimes \sigma_{\pm}^{(j)} \otimes \sigma_z^{\otimes (N-j)} \otimes I^{\otimes (i-1)} \otimes \sigma_{\pm}^{(i)} \otimes \sigma_z^{\otimes (N-i)},\nonumber
\end{eqnarray}
\end{widetext}
Eqs.~(\ref{eq:Lijsigma}) provide the recipe to construct Lindbladian operators for fermionic systems,
based on fermion-to-qubit mapping.
In order to apply Eqs.~(\ref{eq:Lijsigma}), it is enough to specify $N$,
the total number of sites or levels being considered in the system.
Because the expressions are explicit tensorial products, they make numerical implementations
very straightforward.

\section{Quantum dynamics on coupled quantum molecules}
\label{sec:application}
In this section, we proceed to apply the formalism in the context of charged quantum dots~\cite{Kobayashi95,Tarucha96}. In this physical setup, metallic gates are used to confine electrons within a small region of AlGaAs-GaAs~\cite{Fujisawa98}. Manipulating the chemical potential of sources and drains, both described as electronic reservoirs, charges can be introduced inside the nanoestructure~\cite{Oosterkamp98}. Coherent tunneling between two adjacent quantum dots, which form an artificial molecule, permits the codification of information in a qubit, once it is possible to define a two-level system~\cite{Hayashi03}. From the point of view of quantum information processing, two-qubit operations are necessary for the implementation of a universal set of quantum gates. The Coulomb interaction between charges in two molecules~\cite{Shinkai09,Shinkai07} provides a rich dynamics, enough to implement quantum gates~\cite{Fujisawa11}, and the creation of maximally entangled states~\cite{Fanchini10,Oliveira15}.

\begin{figure}[tb]
\centering\includegraphics[width=0.7\linewidth]{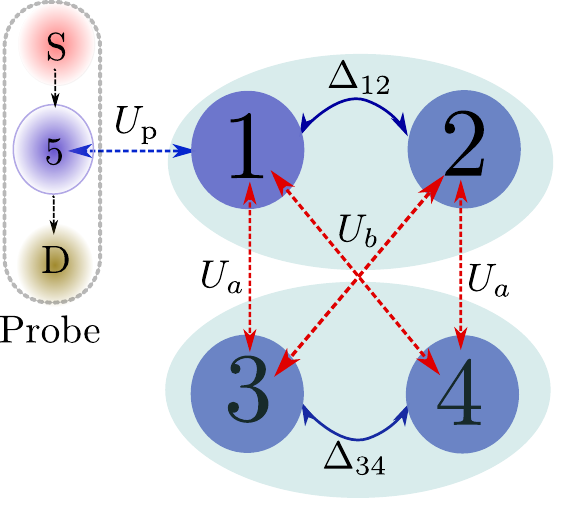}
\caption{Scheme of the coupled quantum molecules with a probe: the first (second) molecule is composed by dots $1$ and $2$ ($3$ and $4$). Tunneling is allowed inside the molecules, described by parameter $\Delta_{ij}$. The electrostatic interactions between electrons inside the molecules, with parameter $U_{a(b)}$, provide the coupling between the molecules. The probe consists of a narrow channel provided by dot $5$, attached to a source (S) and a drain (D), coupled capacitive with $U_{\mathrm{p}}$ as illustrated.}
\label{fig:systemapp}
\end{figure}

We are interested on analyze the quantum dynamics of the coupled quantum molecules, under the effect of a small open quantum system. The last consists on an extra quantum dot with a source and a drain which can be used to charge or discharge the nanoestructure. The configuration of the complete physical system is shown in Fig.~\ref{fig:systemapp}. When a charge occupies the electronic level of this fifth dot, the electrostatic interaction between the charge and the electron in the molecule works as a capacitive probe. The tunneling between probe and molecules is forbidden, so there is no loss of electronic population of the molecule.

Using the formalism of Sec.~\ref{sec:genform}, it is straightforward to obtain the $16$D complete computational basis for the two quantum molecules system, denoted as $\mathcal{M}$. The basis is composed not only of the two-qubit subspace
$\mathcal{M}_{2\mathrm{QB}}=\left\{\ket{1010},\ket{0101},\ket{1001},\ket{0110}\right\}$, but also of others like
the one-particle subspace
$\mathcal{M}_1=\left\{\ket{1000},\ket{0100},\ket{0010},\ket{0001}\right\}$,
the two-particles per molecule subspace
$\mathcal{M}_2=\left\{\ket{1100},\ket{0011}\right\}$,
the three-particle subspace $\mathcal{M}_3=\left\{\ket{1110},\ket{1101},\ket{0111},\ket{1011}\right\}$,
the four-particle state $\ket{1111}$, and the vacuum $\ket{0000}$. The basis allows the description of closed and open quantum dynamics, as the effect of sources, which take an initial vacuum state to some occupied state (eventually the full-occupation state, $|1111\rangle$), or drains, that take any occupied state to the vacuum state, $|0000\rangle$.

For this specific application, we consider that a previous process of initialization prepared the two-qubit system in one of the four state of subspace $\mathcal{M}_{2\mathrm{QB}}$. The Hamiltonian for the two qubits on the coupled quantum molecules is written as
\begin{eqnarray}
H_\mathcal{M}&=&H_0+H_T+H_U\nonumber\\
&=&\bigoplus_{i=1}^4 \varepsilon_i^{(\mathcal{M})}P^{(i)}_++H_T+H_U
\end{eqnarray}
where
\begin{eqnarray}
\label{eq:hsystem}
H_T&=&-\Delta_{12}\left[\sigma^{(1)}_-\otimes\sigma^{(2)}_+\otimes I^{\otimes 2}+\sigma^{(1)}_+\otimes\sigma^{(2)}_-\otimes I^{\otimes 2}\right]\nonumber\\
&-&\Delta_{34}\left[I^{\otimes 2}\otimes\sigma^{(3)}_{-}\otimes\sigma^{(4)}_{+}+I^{\otimes 2}\otimes\sigma^{(3)}_{+}\otimes\sigma^{(4)}_{-}\right],\nonumber\\
H_U&=&U_a\left[P^{(1)}_{+}\otimes I \otimes P^{(3)}_{+} \otimes I + I \otimes P^{(2)}_{+} \otimes I \otimes P^{(4)}_{+}\right]\nonumber \\
&+&U_b\left[P^{(1)}_{+} \otimes I^{\otimes 2}\otimes P^{(4)}_{+}+I \otimes P^{(2)}_{+} \otimes P^{(3)}_{+} \otimes I\right], \nonumber
\end{eqnarray}
with $P_{+}=\sigma_{+}\sigma_{-}=\ket{+}\bra{+}$ and the index $1$ to $4$ describes a specific dot on quantum molecules. The term $H_0$ is the uncoupled term with the electronic energies $\varepsilon_i^{(\mathcal{M})}$, while $H_{T}$ and $H_{U}$ terms describe electronic tunneling inside each molecule and the Coulomb interactions between electrons, respectively.

When the probe is introduced, the full Hamiltonian reads as
\begin{equation}
\label{eq:hfull}
H_{\mathrm{full}}=H_\mathcal{S}\otimes I+I^{\otimes 4}\otimes \varepsilon_5P^{(5)}_+ +U_{\mathrm{p}}P^{(1)}_+\otimes I^{\otimes 3}\otimes P^{(5)}_+,
\end{equation}
where the last term describes the capacitive coupling between system and the extra dot with parameter $U_{\mathrm{p}}$.
To describe the action of both, source and drain of charge on the $5$-th dot,  we use Eq.~(\ref{eq:Lijsigma}) to obtain the Lindblad term
\begin{eqnarray}
 \mathcal{L}_{55}^{\pm} = && I^{\otimes 9} \otimes (\sigma_{\mp} \sigma_{\pm})^{(5)} + I^{\otimes 4} \otimes (\sigma_{\mp} \sigma_{\pm})^{(5)} \otimes I^{\otimes 5}  \nonumber \\
                            && -2 I^{\otimes 4} \otimes \sigma_{\pm}^{(5)} \otimes I^{\otimes 4} \otimes \sigma_{\pm}^{(5)}.
\label{eq:lindblad5dot}
\end{eqnarray}

Our goal is to check the quantum dynamics at the specific condition for generation of maximally entangled states~\cite{Oliveira15}. We start solving numerically Eq.~(\ref{eq:drhodt}), considering the terms of Eq.~(\ref{eq:lindblad5dot}). The solution, $\rho_{\mathcal{S}_5}(t)$, describes the dynamics of system and probe. Then, by tracing out the $5$-th dot, the behavior of the two-qubit system $\mathcal{S}$ (dots $1$ to $4$) is described by the reduced density matrix,
\begin{equation}
 \rho_{\mathcal{M}}(t)=\mathrm{Tr}_{5} [\rho_{\mathcal{S}_5}(t)].
\end{equation}
The average occupation of $i$-th dot ($i=1...4$) is given by
\begin{equation}
 \langle \hat{N}_i \rangle = \mathrm{Tr} \{ \hat{N}_i  \rho_{\mathcal{M}}(t) \},
\end{equation}
where the operator $\hat{N}_i$ is defined as
\begin{equation}
 \hat{N}_i=I^{\otimes i-1} \otimes P_+^{(i)} \otimes I^{\otimes 4-i}.
\end{equation}
To quantify the probabilities of occupation for each state on the two-qubit subspace $\mathcal{M}_{2\mathrm{QB}}$, we calculate the cross-population averages defined as,
\begin{eqnarray}
\label{eq:populations2qb}
P_{\ket{1001}}&=&\langle \hat{N}_1 (I^{\otimes 4}-\hat{N}_2) (I^{\otimes 4}-\hat{N}_3) \hat{N}_4 \rangle,\\
P_{\ket{0110}}&=&\langle (I^{\otimes 4}-\hat{N}_1) \hat{N}_2 \hat{N}_3 (I^{\otimes 4}-\hat{N}_4) \rangle,\nonumber\\
P_{\ket{1010}}&=&\langle \hat{N}_1 (I^{\otimes 4}-\hat{N}_2) \hat{N}_3 (I^{\otimes 4}-\hat{N}_4) \rangle\nonumber\\
P_{\ket{0101}}&=&\langle (I^{\otimes 4}-\hat{N}_1) \hat{N}_2 (I^{\otimes 4}-\hat{N}_3) \hat{N}_4 \rangle\nonumber.
\end{eqnarray}

\begin{figure}[tb]
\centering\includegraphics[width=1\linewidth]{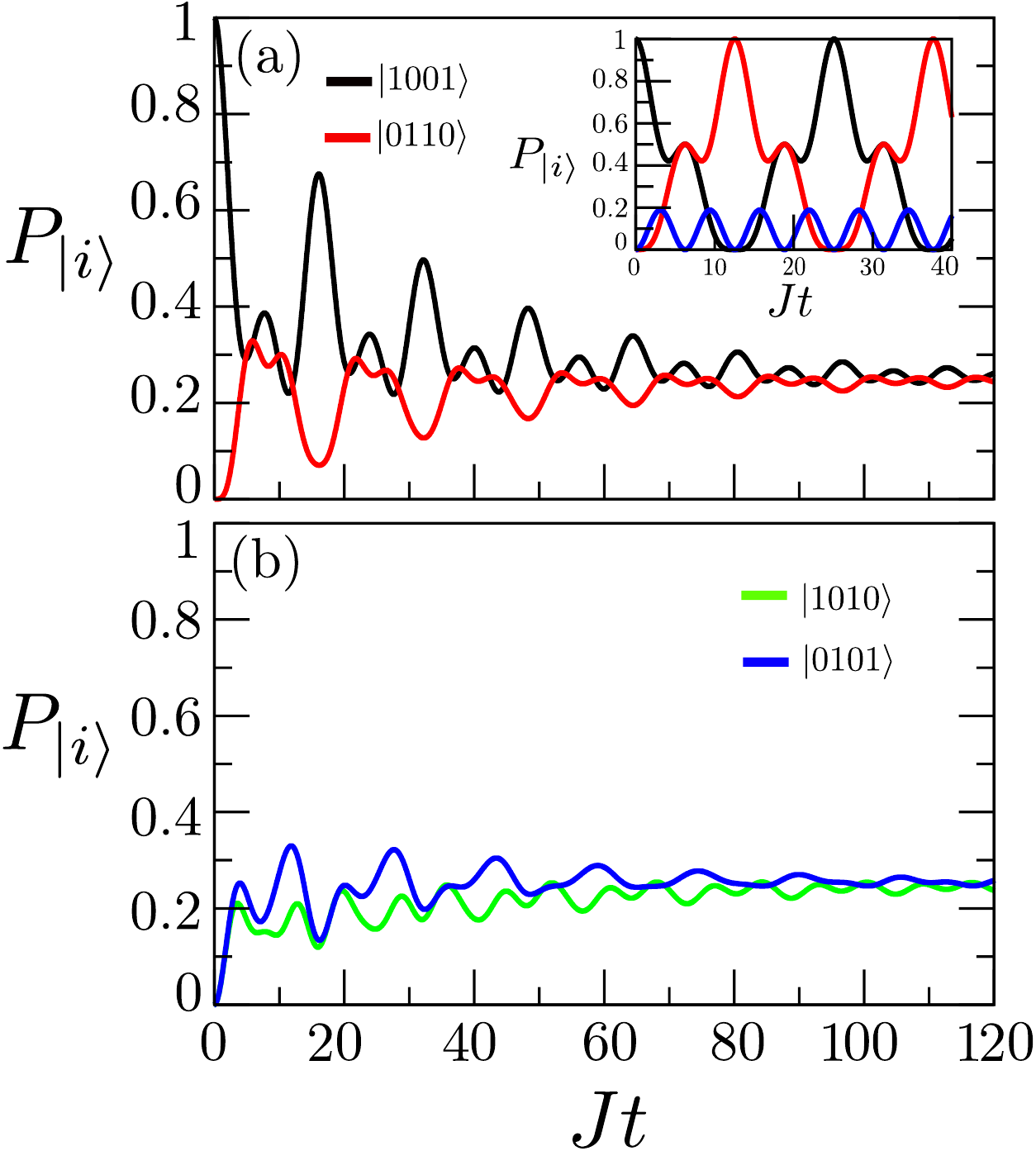}
\caption{Time evolution of cross-populations, Eqs.~(\ref{eq:populations2qb}), for the states belonging to the subspace $\mathcal{M}_{2QB}$, considering probe coupling $U_{\mathrm{p}}=3J$ and $\Gamma_{55}^{(5)}=J$. (a) $P_{\ket{1001}}$ (black line) and $P_{\ket{0110}}$ (red line); (b) $P_{\ket{1010}}$ (green line) and $P_{\ket{0101}}$ (blue line). Inset: Time evolution of all cross-populations (same color choices) when the probe is turned off ($U_{\mathrm{p}}=0$).  Physical parameters are those for creation of Bell states: $\Delta_{12}=\Delta_{34}=\Delta=\sqrt{3}/8 J$, $\varepsilon_{i}^{(\mathcal{M})}=0$, $U_a=(3/4)J$, $U_b=(1/4)J$. }
\label{fig:crosspop}
\end{figure}

As initial condition, we assume that the initialization process prepares the state $\ket{\Psi(0)}=|1001\rangle$ so $P_{|1001\rangle}(0)=1$ ($\langle \hat{N}_1 \rangle = \langle \hat{N}_4 \rangle = 1$). The effect of the probe over the evolution of cross-populations, Eqs.~(\ref{eq:populations2qb}), is shown in Fig.~\ref{fig:crosspop}, considering the physical parameters used to obtain maximally entangled states~\cite{Oliveira15}. For the sake of comparison, we include in the inset the evolution of the same quantities when the probe is turned off, considering a shorter time scale. The periodic coherent dynamics from Ref.~\cite{Oliveira15} is recovered from our approach, where a Bell state is dynamically generate at the times when $P_{|1001\rangle}=P_{|0110\rangle}=0.5$, while $P_{|1010\rangle}=P_{|0101\rangle}=0$.

When the probe is turned on, as a current passes through the narrow conduction channel in dot 5, the effect is to induce an attenuation of the coherent oscillations of populations $P_{|1001\rangle}$ and $P_{|0110\rangle}$, black and red lines on Fig.~\ref{fig:crosspop}(a) respectively. Additionally, we note the increase of populations $P_{|1010\rangle}$ and $P_{|0101\rangle}$, as can be seen from green and blue lines on Fig.~\ref{fig:crosspop}(b). Calculations of the value of population for states in subspaces of $\mathcal{M}$ different from $\mathcal{M}_{2\mathrm{QB}}$ shows that the charges remains confined at the quantum molecules, as expected.  At long times, the population for each state on subspace $\mathcal{M}_{2\mathrm{QB}}$ tend to $0.25$ in the stationary regime.

\begin{figure}[tb]
\centering\includegraphics[width=1\linewidth]{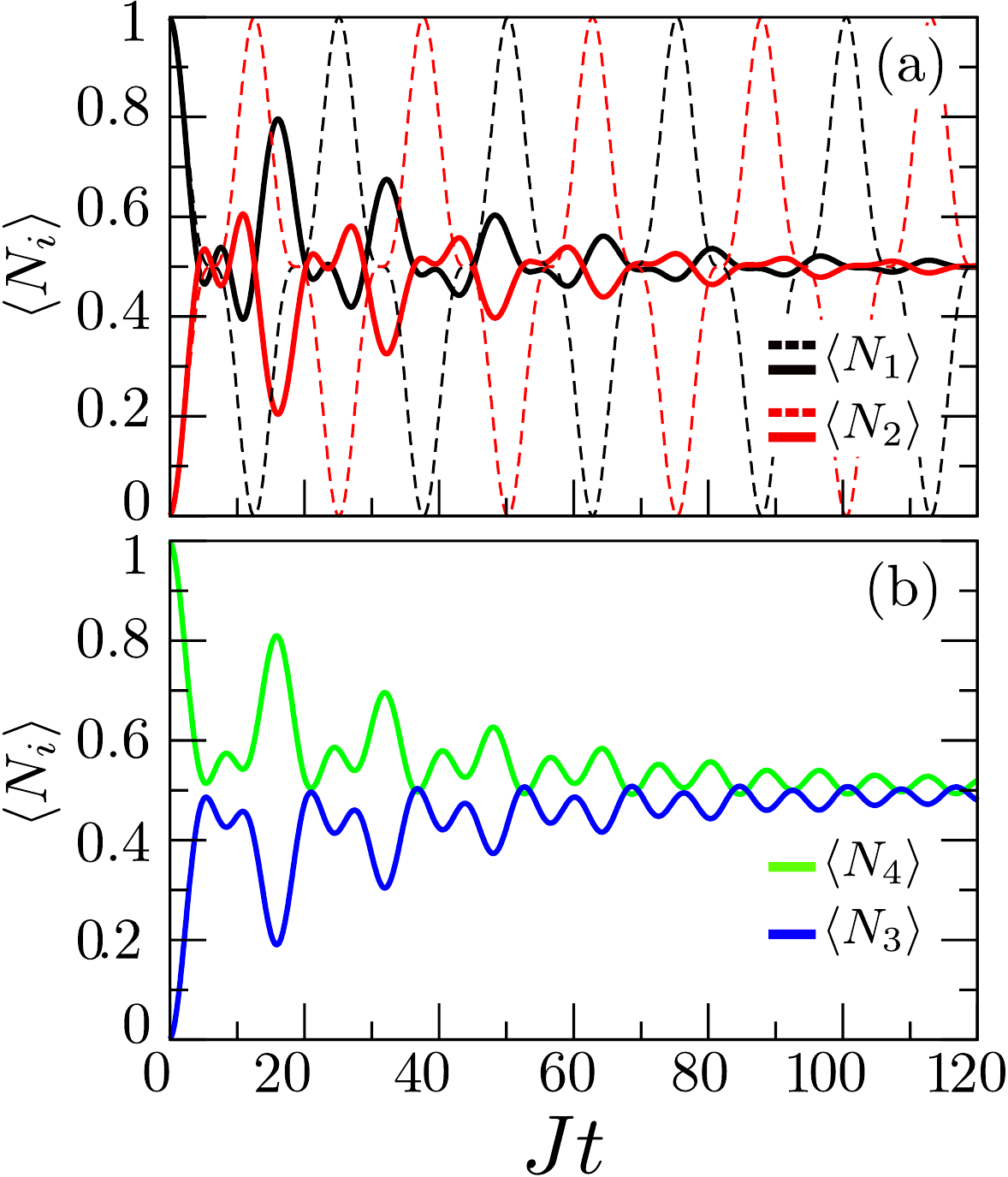}
\caption{Time evolution of the populations of the quantum dots in coupled quantum molecules, for the same physical parameters on Fig~\ref{fig:crosspop}: (a) $\langle \hat{N}_1 \rangle$ (black line) and $\langle \hat{N}_2 \rangle$ (red line), corresponding to the molecule coupled with the probe. We also plot the behavior of both, $\langle \hat{N}_1 \rangle$ (dashed black line) and $\langle \hat{N}_2 \rangle$ (dashed red line) when the probe is turned off ($U_{\mathrm{p}}=0$). (b) $\langle \hat{N}_3 \rangle$ (blue line) and $\langle \hat{N}_4 \rangle$ (green line), corresponding to the second molecule.}
\label{fig:moleculepop}
\end{figure}

To check the dynamics of populations inside each molecule, we calculate the occupations for the single $i$-th quantum dot, $\langle \hat{N}_i \rangle$.
The results for the first molecule (dots $1$ and $2$) are shown in Fig.~\ref{fig:moleculepop}(a), for the same initial condition and physical parameters of Fig.~\ref{fig:crosspop}. If probe is turned off, the occupations $\langle \hat{N}_1 \rangle$ and $\langle \hat{N}_2 \rangle$, shown with dashed lines, develop periodic population inversions, which is a signature of the coherent tunneling between the dots inside the molecule. The same behavior, not shown here, is obtained for the second molecule (dots $3$ and $4$). Once the probe is turned on, the coherent dynamics of $\langle \hat{N}_i \rangle$ is attenuated, as observed in Fig.~\ref{fig:crosspop} for cross-populations. It becomes clear that the second molecule is less affected by the probe, once the oscillations of $\langle N_3\rangle$ and $\langle N_4\rangle$  survive longer than those for the first molecule.

The exact nature of the asymptotic state is the question that rises from the analysis presented above. The occupations of the dots are not able to distinguish between quantum superpositions and mixed states, so it is convenient to analyze the quantum dynamics using the tools of quantum information. Because the physical conditions used for our calculations are the same for the dynamical generation of Bell states, it is interesting to check the behavior of the degree of entanglement in the system. In order to fulfill this task, we use the concurrence as defined by Wooters~\cite{Wootters98}, which is a measurement of entanglement degree between two-qubits. Considering a generic density matrix in a two qubit space $\rho_{\mathrm{2QB}}$, an auxiliary Hermitian operator~\cite{Hill97} $R$ is defined as
\begin{equation}
R=\sqrt{\sqrt{\rho_{\mathrm{2QB}}}\;\widetilde{\rho}_{\mathrm{2QB}}\sqrt{\rho_{\mathrm{2QB}}}},
\end{equation}
where $\widetilde{\rho}_{\mathrm{2QB}}=(\sigma_y \otimes \sigma_y)\rho_{\mathrm{2QB}}^\star (\sigma_y \otimes \sigma_y)$, is the spin-flipped matrix with $\rho_{\mathrm{2QB}}^\star$ being the complex conjugate of $\rho_{\mathrm{2QB}}$.  The concurrence is written as
\begin{equation}
 C=\mathrm{max}(0,\lambda_1-\lambda_2-\lambda_3-\lambda_4),
\end{equation}
where $\lambda_i$ are the eigenvalues of the operator $R$ in decreasing order. For our application, we construct a $4\times 4$ density operator using only the terms on $\rho_{\mathcal{M}}$ related with states of subspace $\mathcal{M}_{\mathrm{2QB}}$. This can be done once the dynamics keeps the other state of the complete basis empty.

Because it is our interest to establish the purity of the coupled molecules, we calculate the linear entropy of the evolved density matrix $\rho_{\mathcal{M}}$, which is defined as
\begin{equation}
S(t)=1-\mathrm{Tr}\left[\rho_{\mathcal{M}}^{2}(t)\right].
\label{eq:linearentro}
\end{equation}
For bipartite quantum systems, the linear entropy works as an entanglement quantifier. Nevertheless, if the quantum system of interest is coupled with an open system, the linear entropy acts as a measurement of purity. If the state of the quantum system remains as a pure quantum state, the linear entropy value is $S=0$. If the quantum system goes to any kind of mixed state $S\neq 0$ with a maximum value $S_{\mathrm{max}}$ given by the expression
\begin{equation}
S_{\mathrm{max}}=1-\frac{1}{d},
\end{equation}
where $d$ is the dimension of the Hilbert space of the quantum system. This value is associated with the statistical mixture of all elements of the basis.

\begin{figure}[tb]
\centering\includegraphics[width=1\linewidth]{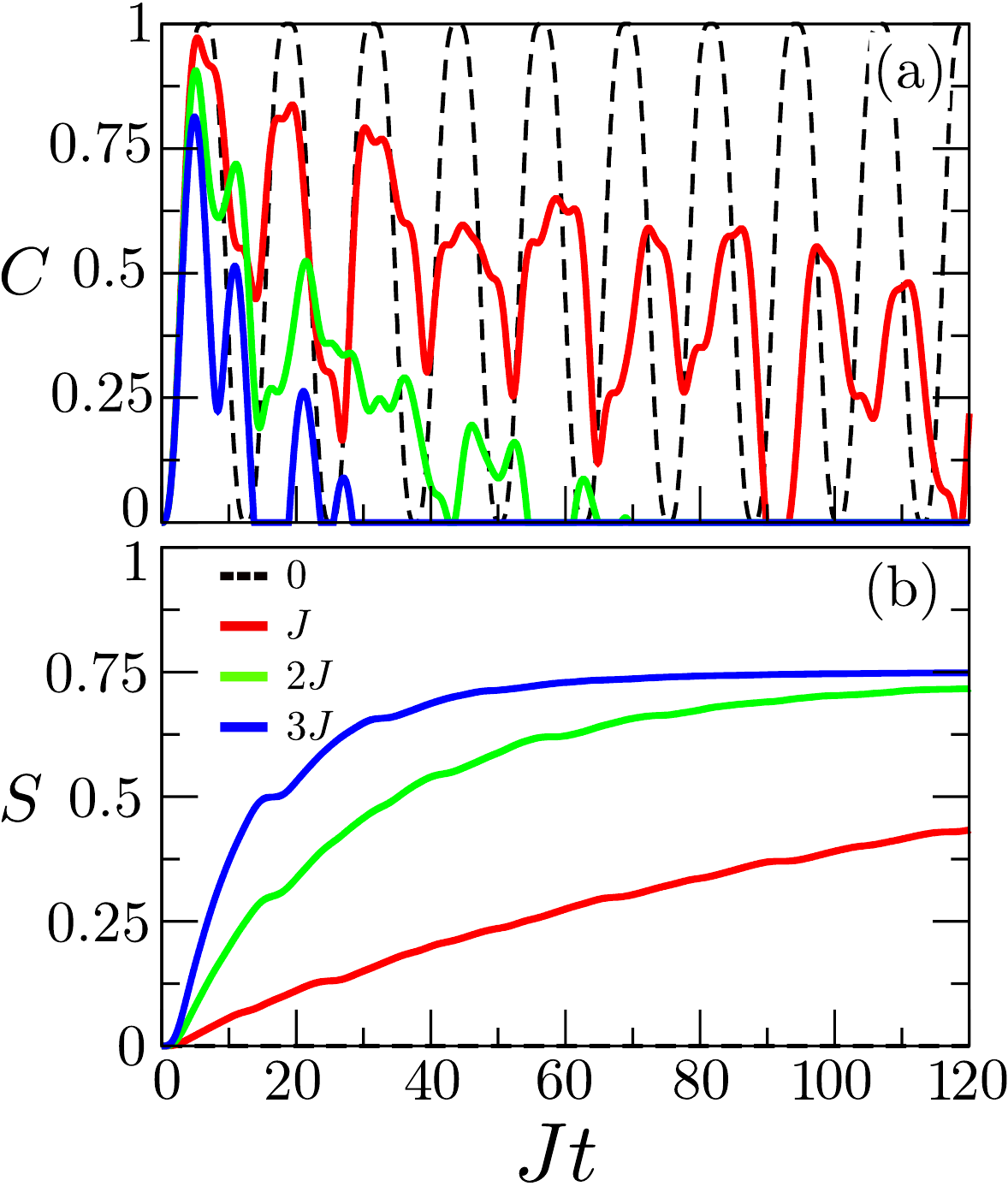}
\caption{Dynamics of (a) concurrence ($C$) and (b) linear entropy ($S$) for several values of coupling $U_{\mathrm{p}}$, for the same choice of physical parameters used on Fig.\ref{fig:crosspop}: $U_{\mathrm{p}}=0$ (black dashed lines), $U_{\mathrm{p}}=J$ (red lines), $U_{\mathrm{p}}=2J$ (green lines) and $U_{\mathrm{p}}=3J$ (blue lines).}
\label{fig:concentro}
\end{figure}

Both quantities are shown in Fig.~\ref{fig:concentro}. Dashed lines in Fig.~\ref{fig:concentro} illustrate the case when the probe is turned off ($U_{\mathrm{p}}=0$). The concurrence, Fig.~\ref{fig:concentro}(a), shows its periodic evolution from a separable ($C=0$) to an entangled Bell state ($C=1$), as discussed in Ref.~\cite{Oliveira15}. For all evolution, the linear entropy in Fig.~\ref{fig:concentro}(b) has a value $S=0$ (dashed line over the $Jt$ axis) which is consistent with the fact that the coupled molecules are a closed system.

This situation changes drastically when the probe is turned on. Red line on Fig.~\ref{fig:concentro} shows a case when the capacitive coupling is comparable with the electrostatic interaction between the electrons inside the molecules so $U_{\mathrm{p}}=J$. The probe changes the concurrence dynamics, Fig.~\ref{fig:concentro}(a), being the main features the lack of periodicity together with the decreasing of the entanglement degree. Around $Jt\sim 90$, occurs a sudden death of entanglement~\cite{Yu06a,Yu06b}, which were demonstrate experimentally in optical setups~\cite{Almeida07} through indirect measurements of concurrence. This phenomenon is an abrupt fall of entanglement to zero, which could be recovered (rebirth) after some time. The purity, red line in Fig.~\ref{fig:concentro}(b), increases smoothly with time, showing that the quantum state inside the molecules becomes a mixed state as times evolves.

As we increase $U_{\mathrm{p}}$, the behavior of concurrence basically shares the same characteristics discussed above with some slight differences. The first is the decrease of temporal scale for the first sudden death and rebirth in Fig.~\ref{fig:concentro}(a). The second is the definitive suppression of the entanglement degree, meaning the asymptotic state has $C=0$. Concerning the purity evolution, green and blue lines in Fig.~\ref{fig:concentro}(b), the time scale to attain the asymptotic mixed state decreases as the coupling parameter $U_{\mathrm{p}}$ increases, revealing the irreversible loss of quantum information on the coupled molecules due to the action of the probe. The results obtained for $U_{\mathrm{p}}=3J$, blue line, confirms the nature of the asymptotical behavior: the value of $S$ is $S_{\mathrm{max}}=1-1/4=0.75$ ($d=4$) which means the system goes to a statistical mixture of all states on $\mathcal{M}_{\mathrm{2QB}}$.

The results presented above characterize a process of decoherence, induced by the action of the probe on the dynamics of the charges inside the coupled quantum molecules. All these aspects together become signatures that the probe induces dephasing on the quantum system without the loss of particles. Additionally, the behavior of both, concurrence and linear entropy, shows that the initial pure system evolves to a statistical mixture when the system is probed.

\section{Summary}
\label{sec:summary}
In this work, we present a formalism for the treatment of the interaction between quantum systems in contact with reservoirs based on a fermion-to-qubit map. The formalism has a flexibility which permits the analysis of general configurations of multipartite systems coupled with multiple reservoirs. We focuss on the obtention of a master equation, where Linblandian operators keep the structure of fermion-to-qubit mapping. The success on the demonstration of such a form of master equation brings all the advantages of Jordan-Wigner transformation to problems of quantum information processing, as used for strong-correlated systems. Specifically, it is possible to treat problem where reservoirs can act as sources and drains of particles. In the particular case of non-interacting reservoirs prepared as thermal states, the method provides expressions for Lindblad super-operators that can be straightforwardly use on numerical implementations of non-equilibrium problems.

To illustrate, we apply our formalism to the problem of dynamics of two electrons inside quantum molecules in contact with a probe, the last being an open system. The probe is a narrow transmission channel, being an open quantum dot attached to source and drain leads. By using the general equation for Lindblad super-operators, we obtain a reduced density matrix for the coupled quantum molecules. Our calculations of populations, concurrence and linear entropy let us to conclude that the probe induces dephasing, which makes the system lose the ability to generate entangled Bell states as time evolves. It is worth to remark an interesting feature induced by the probe: the apparition of sudden deaths and rebirths of entanglement. This sudden death is usually explained as caused by the action of quantum noise over the composite entangled bipartite system. Additionally, the system evolves to an asymptotic state, being a statistical mixture of the four elements of the subspace $\mathcal{M}_{2\mathrm{QB}}=\left\{\ket{1010},\ket{0101},\ket{1001},\ket{0110}\right\}$, indicated by a linear entropy compatible to the number of states in a reduced Hilbert space.
\section{Acknowledgments}
This work was supported by CNPq (grant 307464/2015-6), FAPEMIG (grant APQ-01768-14) and the Brazilian National Institute of Science and Technology of Quantum Information (INCT-IQ).


\begin{thebibliography}{58}%
\makeatletter
\providecommand \@ifxundefined [1]{%
 \@ifx{#1\undefined}
}%
\providecommand \@ifnum [1]{%
 \ifnum #1\expandafter \@firstoftwo
 \else \expandafter \@secondoftwo
 \fi
}%
\providecommand \@ifx [1]{%
 \ifx #1\expandafter \@firstoftwo
 \else \expandafter \@secondoftwo
 \fi
}%
\providecommand \natexlab [1]{#1}%
\providecommand \enquote  [1]{``#1''}%
\providecommand \bibnamefont  [1]{#1}%
\providecommand \bibfnamefont [1]{#1}%
\providecommand \citenamefont [1]{#1}%
\providecommand \href@noop [0]{\@secondoftwo}%
\providecommand \href [0]{\begingroup \@sanitize@url \@href}%
\providecommand \@href[1]{\@@startlink{#1}\@@href}%
\providecommand \@@href[1]{\endgroup#1\@@endlink}%
\providecommand \@sanitize@url [0]{\catcode `\\12\catcode `\$12\catcode
  `\&12\catcode `\#12\catcode `\^12\catcode `\_12\catcode `\%12\relax}%
\providecommand \@@startlink[1]{}%
\providecommand \@@endlink[0]{}%
\providecommand \url  [0]{\begingroup\@sanitize@url \@url }%
\providecommand \@url [1]{\endgroup\@href {#1}{\urlprefix }}%
\providecommand \urlprefix  [0]{URL }%
\providecommand \Eprint [0]{\href }%
\providecommand \doibase [0]{http://dx.doi.org/}%
\providecommand \selectlanguage [0]{\@gobble}%
\providecommand \bibinfo  [0]{\@secondoftwo}%
\providecommand \bibfield  [0]{\@secondoftwo}%
\providecommand \translation [1]{[#1]}%
\providecommand \BibitemOpen [0]{}%
\providecommand \bibitemStop [0]{}%
\providecommand \bibitemNoStop [0]{.\EOS\space}%
\providecommand \EOS [0]{\spacefactor3000\relax}%
\providecommand \BibitemShut  [1]{\csname bibitem#1\endcsname}%
\let\auto@bib@innerbib\@empty
\bibitem [{\citenamefont {Zurek}(1991)}]{Zurek91}%
  \BibitemOpen
  \bibfield  {author} {\bibinfo {author} {\bibfnamefont {W.~H.}\ \bibnamefont
  {Zurek}},\ }\href@noop {} {\bibfield  {journal} {\bibinfo  {journal} {Physics
  Today}\ }\textbf {\bibinfo {volume} {44}},\ \bibinfo {pages} {36} (\bibinfo
  {year} {1991})}\BibitemShut {NoStop}%
\bibitem [{\citenamefont {Haroche}(1998)}]{HarochePT98}%
  \BibitemOpen
  \bibfield  {author} {\bibinfo {author} {\bibfnamefont {S.}~\bibnamefont
  {Haroche}},\ }\href@noop {} {\bibfield  {journal} {\bibinfo  {journal}
  {Physics Today}\ }\textbf {\bibinfo {volume} {51}},\ \bibinfo {pages} {36}
  (\bibinfo {year} {1998})}\BibitemShut {NoStop}%
\bibitem [{\citenamefont {Schlosshauer}(2007)}]{Schlosshauerbook}%
  \BibitemOpen
  \bibfield  {author} {\bibinfo {author} {\bibfnamefont {M.}~\bibnamefont
  {Schlosshauer}},\ }\href@noop {} {\emph {\bibinfo {title} {{Decoherence and
  the Quantum-to-Classical Transition}}}},\ \bibinfo {edition} {1st}\ ed.\
  (\bibinfo  {publisher} {Springer-Verlag},\ \bibinfo {address} {Berlin
  Heidelberg},\ \bibinfo {year} {2007})\BibitemShut {NoStop}%
\bibitem [{\citenamefont {Schlosshauer}(2005)}]{Schlosshauer05}%
  \BibitemOpen
  \bibfield  {author} {\bibinfo {author} {\bibfnamefont {M.}~\bibnamefont
  {Schlosshauer}},\ }\href {\doibase 10.1103/RevModPhys.76.1267} {\bibfield
  {journal} {\bibinfo  {journal} {Rev. Mod. Phys.}\ }\textbf {\bibinfo {volume}
  {76}},\ \bibinfo {pages} {1267} (\bibinfo {year} {2005})}\BibitemShut
  {NoStop}%
\bibitem [{\citenamefont {Schlosshauer}(2014)}]{Schlosshauer2014}%
  \BibitemOpen
  \bibfield  {author} {\bibinfo {author} {\bibfnamefont {M.}~\bibnamefont
  {Schlosshauer}},\ }\href@noop {} {\bibfield  {journal} {\bibinfo  {journal}
  {arXiv:1404.2635 [quant-ph]}\ } (\bibinfo {year} {2014})}\BibitemShut
  {NoStop}%
\bibitem [{\citenamefont {Breuer}\ and\ \citenamefont
  {Petruccione}(2007)}]{Breuerbook}%
  \BibitemOpen
  \bibfield  {author} {\bibinfo {author} {\bibfnamefont {H.-P.}\ \bibnamefont
  {Breuer}}\ and\ \bibinfo {author} {\bibfnamefont {F.}~\bibnamefont
  {Petruccione}},\ }\href@noop {} {\emph {\bibinfo {title} {{The Theory of Open
  Quantum Systems}}}},\ \bibinfo {edition} {1st}\ ed.\ (\bibinfo  {publisher}
  {Oxford University Press},\ \bibinfo {address} {Oxford},\ \bibinfo {year}
  {2007})\BibitemShut {NoStop}%
\bibitem [{\citenamefont {Fano}(1957)}]{Fano57}%
  \BibitemOpen
  \bibfield  {author} {\bibinfo {author} {\bibfnamefont {U.}~\bibnamefont
  {Fano}},\ }\href {\doibase 10.1103/RevModPhys.29.74} {\bibfield  {journal}
  {\bibinfo  {journal} {Rev. Mod. Phys.}\ }\textbf {\bibinfo {volume} {29}},\
  \bibinfo {pages} {74} (\bibinfo {year} {1957})}\BibitemShut {NoStop}%
\bibitem [{\citenamefont {Gorini}\ \emph {et~al.}(1976)\citenamefont {Gorini},
  \citenamefont {Kossakowski},\ and\ \citenamefont {Sudarshan}}]{Gorini76}%
  \BibitemOpen
  \bibfield  {author} {\bibinfo {author} {\bibfnamefont {V.}~\bibnamefont
  {Gorini}}, \bibinfo {author} {\bibfnamefont {A.}~\bibnamefont {Kossakowski}},
  \ and\ \bibinfo {author} {\bibfnamefont {E.~C.~G.}\ \bibnamefont
  {Sudarshan}},\ }\href@noop {} {\bibfield  {journal} {\bibinfo  {journal} {J.
  Math. Phys.}\ }\textbf {\bibinfo {volume} {71}},\ \bibinfo {pages} {821}
  (\bibinfo {year} {1976})}\BibitemShut {NoStop}%
\bibitem [{\citenamefont {Lindblad}(1976)}]{Lindblad76}%
  \BibitemOpen
  \bibfield  {author} {\bibinfo {author} {\bibfnamefont {G.}~\bibnamefont
  {Lindblad}},\ }\href@noop {} {\bibfield  {journal} {\bibinfo  {journal}
  {Commun. Math. Phys.}\ }\textbf {\bibinfo {volume} {48}},\ \bibinfo {pages}
  {119} (\bibinfo {year} {1976})}\BibitemShut {NoStop}%
\bibitem [{\citenamefont {Zhang}\ \emph
  {et~al.}(2012{\natexlab{a}})\citenamefont {Zhang}, \citenamefont {Lo},
  \citenamefont {Xiong}, \citenamefont {Tu},\ and\ \citenamefont
  {Nori}}]{zhang2012general}%
  \BibitemOpen
  \bibfield  {author} {\bibinfo {author} {\bibfnamefont {W.-M.}\ \bibnamefont
  {Zhang}}, \bibinfo {author} {\bibfnamefont {P.-Y.}\ \bibnamefont {Lo}},
  \bibinfo {author} {\bibfnamefont {H.-N.}\ \bibnamefont {Xiong}}, \bibinfo
  {author} {\bibfnamefont {M.~W.-Y.}\ \bibnamefont {Tu}}, \ and\ \bibinfo
  {author} {\bibfnamefont {F.}~\bibnamefont {Nori}},\ }\href@noop {} {\bibfield
   {journal} {\bibinfo  {journal} {Phys. Rev. Lett.}\ }\textbf {\bibinfo
  {volume} {109}},\ \bibinfo {pages} {170402} (\bibinfo {year}
  {2012}{\natexlab{a}})}\BibitemShut {NoStop}%
\bibitem [{\citenamefont {Breuer}\ \emph {et~al.}(2016)\citenamefont {Breuer},
  \citenamefont {Laine}, \citenamefont {Piilo},\ and\ \citenamefont
  {Vacchini}}]{BreuerRMP16}%
  \BibitemOpen
  \bibfield  {author} {\bibinfo {author} {\bibfnamefont {H.-P.}\ \bibnamefont
  {Breuer}}, \bibinfo {author} {\bibfnamefont {E.-M.}\ \bibnamefont {Laine}},
  \bibinfo {author} {\bibfnamefont {J.}~\bibnamefont {Piilo}}, \ and\ \bibinfo
  {author} {\bibfnamefont {B.}~\bibnamefont {Vacchini}},\ }\href {\doibase
  10.1103/RevModPhys.88.021002} {\bibfield  {journal} {\bibinfo  {journal}
  {Rev. Mod. Phys.}\ }\textbf {\bibinfo {volume} {88}},\ \bibinfo {pages}
  {021002} (\bibinfo {year} {2016})}\BibitemShut {NoStop}%
\bibitem [{\citenamefont {Liuzzo-Scorpo}\ \emph {et~al.}(2017)\citenamefont
  {Liuzzo-Scorpo}, \citenamefont {Roga}, \citenamefont {Souza}, \citenamefont
  {Bernardes},\ and\ \citenamefont {Adesso}}]{Scorpo17}%
  \BibitemOpen
  \bibfield  {author} {\bibinfo {author} {\bibfnamefont {P.}~\bibnamefont
  {Liuzzo-Scorpo}}, \bibinfo {author} {\bibfnamefont {W.}~\bibnamefont {Roga}},
  \bibinfo {author} {\bibfnamefont {L.~A.~M.}\ \bibnamefont {Souza}}, \bibinfo
  {author} {\bibfnamefont {N.~K.}\ \bibnamefont {Bernardes}}, \ and\ \bibinfo
  {author} {\bibfnamefont {G.}~\bibnamefont {Adesso}},\ }\href {\doibase
  10.1103/PhysRevLett.118.050401} {\bibfield  {journal} {\bibinfo  {journal}
  {Phys. Rev. Lett.}\ }\textbf {\bibinfo {volume} {118}},\ \bibinfo {pages}
  {050401} (\bibinfo {year} {2017})}\BibitemShut {NoStop}%
\bibitem [{\citenamefont {Bernardes}\ \emph {et~al.}(2017)\citenamefont
  {Bernardes}, \citenamefont {Carvalho}, \citenamefont {Monken},\ and\
  \citenamefont {Santos}}]{Bernardes2017}%
  \BibitemOpen
  \bibfield  {author} {\bibinfo {author} {\bibfnamefont {N.~K.}\ \bibnamefont
  {Bernardes}}, \bibinfo {author} {\bibfnamefont {A.~R.~R.}\ \bibnamefont
  {Carvalho}}, \bibinfo {author} {\bibfnamefont {C.~H.}\ \bibnamefont
  {Monken}}, \ and\ \bibinfo {author} {\bibfnamefont {M.~F.}\ \bibnamefont
  {Santos}},\ }\href@noop {} {\bibfield  {journal} {\bibinfo  {journal} {Phys.
  Rev. A}\ }\textbf {\bibinfo {volume} {95}},\ \bibinfo {pages} {32117}
  (\bibinfo {year} {2017})}\BibitemShut {NoStop}%
\bibitem [{\citenamefont {Li}\ \emph {et~al.}(2011)\citenamefont {Li},
  \citenamefont {Tang}, \citenamefont {Li},\ and\ \citenamefont {Guo}}]{Li11}%
  \BibitemOpen
  \bibfield  {author} {\bibinfo {author} {\bibfnamefont {C.-F.}\ \bibnamefont
  {Li}}, \bibinfo {author} {\bibfnamefont {J.-S.}\ \bibnamefont {Tang}},
  \bibinfo {author} {\bibfnamefont {Y.-L.}\ \bibnamefont {Li}}, \ and\ \bibinfo
  {author} {\bibfnamefont {G.-C.}\ \bibnamefont {Guo}},\ }\href {\doibase
  10.1103/PhysRevA.83.064102} {\bibfield  {journal} {\bibinfo  {journal} {Phys.
  Rev. A}\ }\textbf {\bibinfo {volume} {83}},\ \bibinfo {pages} {064102}
  (\bibinfo {year} {2011})}\BibitemShut {NoStop}%
\bibitem [{\citenamefont {Liu}\ \emph {et~al.}(2011)\citenamefont {Liu},
  \citenamefont {Li}, \citenamefont {Huang}, \citenamefont {Li}, \citenamefont
  {Guo}, \citenamefont {Laine}, \citenamefont {Breuer},\ and\ \citenamefont
  {Piilo}}]{liu2011experimental}%
  \BibitemOpen
  \bibfield  {author} {\bibinfo {author} {\bibfnamefont {B.-H.}\ \bibnamefont
  {Liu}}, \bibinfo {author} {\bibfnamefont {L.}~\bibnamefont {Li}}, \bibinfo
  {author} {\bibfnamefont {Y.-F.}\ \bibnamefont {Huang}}, \bibinfo {author}
  {\bibfnamefont {C.-F.}\ \bibnamefont {Li}}, \bibinfo {author} {\bibfnamefont
  {G.-C.}\ \bibnamefont {Guo}}, \bibinfo {author} {\bibfnamefont {E.-M.}\
  \bibnamefont {Laine}}, \bibinfo {author} {\bibfnamefont {H.-P.}\ \bibnamefont
  {Breuer}}, \ and\ \bibinfo {author} {\bibfnamefont {J.}~\bibnamefont
  {Piilo}},\ }\href@noop {} {\bibfield  {journal} {\bibinfo  {journal} {Nat.
  Phys.}\ }\textbf {\bibinfo {volume} {7}},\ \bibinfo {pages} {931} (\bibinfo
  {year} {2011})}\BibitemShut {NoStop}%
\bibitem [{\citenamefont {Jauho}\ \emph {et~al.}(1994)\citenamefont {Jauho},
  \citenamefont {Wingreen},\ and\ \citenamefont {Meir}}]{Jauho94}%
  \BibitemOpen
  \bibfield  {author} {\bibinfo {author} {\bibfnamefont {A.-P.}\ \bibnamefont
  {Jauho}}, \bibinfo {author} {\bibfnamefont {N.~S.}\ \bibnamefont {Wingreen}},
  \ and\ \bibinfo {author} {\bibfnamefont {Y.}~\bibnamefont {Meir}},\ }\href
  {\doibase 10.1103/PhysRevB.50.5528} {\bibfield  {journal} {\bibinfo
  {journal} {Phys. Rev. B}\ }\textbf {\bibinfo {volume} {50}},\ \bibinfo
  {pages} {5528} (\bibinfo {year} {1994})}\BibitemShut {NoStop}%
\bibitem [{\citenamefont {Platero}\ and\ \citenamefont
  {Aguado}(2004)}]{platero2004photon}%
  \BibitemOpen
  \bibfield  {author} {\bibinfo {author} {\bibfnamefont {G.}~\bibnamefont
  {Platero}}\ and\ \bibinfo {author} {\bibfnamefont {R.}~\bibnamefont
  {Aguado}},\ }\href@noop {} {\bibfield  {journal} {\bibinfo  {journal} {Phys.
  Rep.}\ }\textbf {\bibinfo {volume} {395}},\ \bibinfo {pages} {1} (\bibinfo
  {year} {2004})}\BibitemShut {NoStop}%
\bibitem [{\citenamefont {Cota}\ \emph {et~al.}(2005)\citenamefont {Cota},
  \citenamefont {Aguado},\ and\ \citenamefont {Platero}}]{cota2005ac}%
  \BibitemOpen
  \bibfield  {author} {\bibinfo {author} {\bibfnamefont {E.}~\bibnamefont
  {Cota}}, \bibinfo {author} {\bibfnamefont {R.}~\bibnamefont {Aguado}}, \ and\
  \bibinfo {author} {\bibfnamefont {G.}~\bibnamefont {Platero}},\ }\href@noop
  {} {\bibfield  {journal} {\bibinfo  {journal} {Phys. Rev. Lett.}\ }\textbf
  {\bibinfo {volume} {94}},\ \bibinfo {pages} {107202} (\bibinfo {year}
  {2005})}\BibitemShut {NoStop}%
\bibitem [{\citenamefont {Souza}(2007)}]{souza2007spin}%
  \BibitemOpen
  \bibfield  {author} {\bibinfo {author} {\bibfnamefont {F.~M.}\ \bibnamefont
  {Souza}},\ }\href@noop {} {\bibfield  {journal} {\bibinfo  {journal} {Phys.
  Rev. B}\ }\textbf {\bibinfo {volume} {76}},\ \bibinfo {pages} {205315}
  (\bibinfo {year} {2007})}\BibitemShut {NoStop}%
\bibitem [{\citenamefont {Souza}\ \emph {et~al.}(2007)\citenamefont {Souza},
  \citenamefont {Le\~ao}, \citenamefont {Gester},\ and\ \citenamefont
  {Jauho}}]{souza2007transient}%
  \BibitemOpen
  \bibfield  {author} {\bibinfo {author} {\bibfnamefont {F.~M.}\ \bibnamefont
  {Souza}}, \bibinfo {author} {\bibfnamefont {S.~A.}\ \bibnamefont {Le\~ao}},
  \bibinfo {author} {\bibfnamefont {R.~M.}\ \bibnamefont {Gester}}, \ and\
  \bibinfo {author} {\bibfnamefont {A.-P.}\ \bibnamefont {Jauho}},\ }\href@noop
  {} {\bibfield  {journal} {\bibinfo  {journal} {Phys. Rev. B}\ }\textbf
  {\bibinfo {volume} {76}},\ \bibinfo {pages} {125318} (\bibinfo {year}
  {2007})}\BibitemShut {NoStop}%
\bibitem [{\citenamefont {Trocha}(2010)}]{trocha2010beating}%
  \BibitemOpen
  \bibfield  {author} {\bibinfo {author} {\bibfnamefont {P.}~\bibnamefont
  {Trocha}},\ }\href@noop {} {\bibfield  {journal} {\bibinfo  {journal} {Phys.
  Rev. B}\ }\textbf {\bibinfo {volume} {82}},\ \bibinfo {pages} {115320}
  (\bibinfo {year} {2010})}\BibitemShut {NoStop}%
\bibitem [{\citenamefont {Perfetto}\ \emph {et~al.}(2010)\citenamefont
  {Perfetto}, \citenamefont {Stefanucci},\ and\ \citenamefont
  {Cini}}]{perfetto2010correlation}%
  \BibitemOpen
  \bibfield  {author} {\bibinfo {author} {\bibfnamefont {E.}~\bibnamefont
  {Perfetto}}, \bibinfo {author} {\bibfnamefont {G.}~\bibnamefont
  {Stefanucci}}, \ and\ \bibinfo {author} {\bibfnamefont {M.}~\bibnamefont
  {Cini}},\ }\href@noop {} {\bibfield  {journal} {\bibinfo  {journal} {Phys.
  Rev. Lett.}\ }\textbf {\bibinfo {volume} {105}},\ \bibinfo {pages} {156802}
  (\bibinfo {year} {2010})}\BibitemShut {NoStop}%
\bibitem [{\citenamefont {Assun\c{c}\~{a}o}\ \emph {et~al.}(2013)\citenamefont
  {Assun\c{c}\~{a}o}, \citenamefont {de~Oliveira}, \citenamefont
  {Villas-B\^oas},\ and\ \citenamefont {Souza}}]{Assuncao13}%
  \BibitemOpen
  \bibfield  {author} {\bibinfo {author} {\bibfnamefont {M.~O.}\ \bibnamefont
  {Assun\c{c}\~{a}o}}, \bibinfo {author} {\bibfnamefont {E.~J.~R.}\
  \bibnamefont {de~Oliveira}}, \bibinfo {author} {\bibfnamefont {J.~M.}\
  \bibnamefont {Villas-B\^oas}}, \ and\ \bibinfo {author} {\bibfnamefont
  {F.~M.}\ \bibnamefont {Souza}},\ }\href@noop {} {\bibfield  {journal}
  {\bibinfo  {journal} {J. Phys.: Condens Matter}\ }\textbf {\bibinfo {volume}
  {25}},\ \bibinfo {pages} {135301} (\bibinfo {year} {2013})}\BibitemShut
  {NoStop}%
\bibitem [{\citenamefont {Odashima}\ and\ \citenamefont
  {Lewenkopf}(2017)}]{odashima2017time}%
  \BibitemOpen
  \bibfield  {author} {\bibinfo {author} {\bibfnamefont {M.~M.}\ \bibnamefont
  {Odashima}}\ and\ \bibinfo {author} {\bibfnamefont {C.~H.}\ \bibnamefont
  {Lewenkopf}},\ }\href@noop {} {\bibfield  {journal} {\bibinfo  {journal}
  {Phys. Rev. B}\ }\textbf {\bibinfo {volume} {95}},\ \bibinfo {pages} {104301}
  (\bibinfo {year} {2017})}\BibitemShut {NoStop}%
\bibitem [{\citenamefont {Liu}\ \emph {et~al.}(2014)\citenamefont {Liu},
  \citenamefont {Petersson}, \citenamefont {Stehlik}, \citenamefont {Taylor},\
  and\ \citenamefont {Petta}}]{liu2014photon}%
  \BibitemOpen
  \bibfield  {author} {\bibinfo {author} {\bibfnamefont {Y.-Y.}\ \bibnamefont
  {Liu}}, \bibinfo {author} {\bibfnamefont {K.~D.}\ \bibnamefont {Petersson}},
  \bibinfo {author} {\bibfnamefont {J.}~\bibnamefont {Stehlik}}, \bibinfo
  {author} {\bibfnamefont {J.~M.}\ \bibnamefont {Taylor}}, \ and\ \bibinfo
  {author} {\bibfnamefont {J.~R.}\ \bibnamefont {Petta}},\ }\href@noop {}
  {\bibfield  {journal} {\bibinfo  {journal} {Phys. Rev. Lett.}\ }\textbf
  {\bibinfo {volume} {113}},\ \bibinfo {pages} {036801} (\bibinfo {year}
  {2014})}\BibitemShut {NoStop}%
\bibitem [{\citenamefont {Kulkarni}\ \emph {et~al.}(2014)\citenamefont
  {Kulkarni}, \citenamefont {Cotlet},\ and\ \citenamefont
  {T{\"u}reci}}]{kulkarni2014cavity}%
  \BibitemOpen
  \bibfield  {author} {\bibinfo {author} {\bibfnamefont {M.}~\bibnamefont
  {Kulkarni}}, \bibinfo {author} {\bibfnamefont {O.}~\bibnamefont {Cotlet}}, \
  and\ \bibinfo {author} {\bibfnamefont {H.~E.}\ \bibnamefont {T{\"u}reci}},\
  }\href@noop {} {\bibfield  {journal} {\bibinfo  {journal} {Phys. Rev. B}\
  }\textbf {\bibinfo {volume} {90}},\ \bibinfo {pages} {125402} (\bibinfo
  {year} {2014})}\BibitemShut {NoStop}%
\bibitem [{\citenamefont {H{\"a}rtle}\ and\ \citenamefont
  {Kulkarni}(2015)}]{hartle2015effect}%
  \BibitemOpen
  \bibfield  {author} {\bibinfo {author} {\bibfnamefont {R.}~\bibnamefont
  {H{\"a}rtle}}\ and\ \bibinfo {author} {\bibfnamefont {M.}~\bibnamefont
  {Kulkarni}},\ }\href@noop {} {\bibfield  {journal} {\bibinfo  {journal}
  {Phys. Rev. B}\ }\textbf {\bibinfo {volume} {91}},\ \bibinfo {pages} {245429}
  (\bibinfo {year} {2015})}\BibitemShut {NoStop}%
\bibitem [{\citenamefont {Purkayastha}\ \emph {et~al.}(2016)\citenamefont
  {Purkayastha}, \citenamefont {Dhar},\ and\ \citenamefont
  {Kulkarni}}]{purkayastha2016out}%
  \BibitemOpen
  \bibfield  {author} {\bibinfo {author} {\bibfnamefont {A.}~\bibnamefont
  {Purkayastha}}, \bibinfo {author} {\bibfnamefont {A.}~\bibnamefont {Dhar}}, \
  and\ \bibinfo {author} {\bibfnamefont {M.}~\bibnamefont {Kulkarni}},\
  }\href@noop {} {\bibfield  {journal} {\bibinfo  {journal} {Phys. Rev. A}\
  }\textbf {\bibinfo {volume} {93}},\ \bibinfo {pages} {062114} (\bibinfo
  {year} {2016})}\BibitemShut {NoStop}%
\bibitem [{\citenamefont {Agarwalla}\ \emph {et~al.}(2016)\citenamefont
  {Agarwalla}, \citenamefont {Kulkarni}, \citenamefont {Mukamel},\ and\
  \citenamefont {Segal}}]{agarwalla2016tunable}%
  \BibitemOpen
  \bibfield  {author} {\bibinfo {author} {\bibfnamefont {B.~K.}\ \bibnamefont
  {Agarwalla}}, \bibinfo {author} {\bibfnamefont {M.}~\bibnamefont {Kulkarni}},
  \bibinfo {author} {\bibfnamefont {S.}~\bibnamefont {Mukamel}}, \ and\
  \bibinfo {author} {\bibfnamefont {D.}~\bibnamefont {Segal}},\ }\href@noop {}
  {\bibfield  {journal} {\bibinfo  {journal} {Phys. Rev. B}\ }\textbf {\bibinfo
  {volume} {94}},\ \bibinfo {pages} {035434} (\bibinfo {year}
  {2016})}\BibitemShut {NoStop}%
\bibitem [{\citenamefont {Reichert}\ \emph {et~al.}(2016)\citenamefont
  {Reichert}, \citenamefont {Nalbach},\ and\ \citenamefont
  {Thorwart}}]{reichert2016dynamics}%
  \BibitemOpen
  \bibfield  {author} {\bibinfo {author} {\bibfnamefont {J.}~\bibnamefont
  {Reichert}}, \bibinfo {author} {\bibfnamefont {P.}~\bibnamefont {Nalbach}}, \
  and\ \bibinfo {author} {\bibfnamefont {M.}~\bibnamefont {Thorwart}},\
  }\href@noop {} {\bibfield  {journal} {\bibinfo  {journal} {Phys. Rev. A}\
  }\textbf {\bibinfo {volume} {94}},\ \bibinfo {pages} {032127} (\bibinfo
  {year} {2016})}\BibitemShut {NoStop}%
\bibitem [{\citenamefont {Mann}\ \emph {et~al.}(2016)\citenamefont {Mann},
  \citenamefont {Br{\"u}ggemann},\ and\ \citenamefont
  {Thorwart}}]{mann2016dissipative}%
  \BibitemOpen
  \bibfield  {author} {\bibinfo {author} {\bibfnamefont {N.}~\bibnamefont
  {Mann}}, \bibinfo {author} {\bibfnamefont {J.}~\bibnamefont
  {Br{\"u}ggemann}}, \ and\ \bibinfo {author} {\bibfnamefont {M.}~\bibnamefont
  {Thorwart}},\ }\href@noop {} {\bibfield  {journal} {\bibinfo  {journal} {Eur.
  Phys. J. B}\ }\textbf {\bibinfo {volume} {89}},\ \bibinfo {pages} {279}
  (\bibinfo {year} {2016})}\BibitemShut {NoStop}%
\bibitem [{\citenamefont {Yang}\ and\ \citenamefont {Zhang}(2017)}]{Yang17}%
  \BibitemOpen
  \bibfield  {author} {\bibinfo {author} {\bibfnamefont {P.-Y.}\ \bibnamefont
  {Yang}}\ and\ \bibinfo {author} {\bibfnamefont {W.-M.}\ \bibnamefont
  {Zhang}},\ }\href@noop {} {\bibfield  {journal} {\bibinfo  {journal}
  {Frontiers of Physics}\ }\textbf {\bibinfo {volume} {12}},\ \bibinfo {pages}
  {127204} (\bibinfo {year} {2017})}\BibitemShut {NoStop}%
\bibitem [{\citenamefont {Xiong}\ \emph {et~al.}(2015)\citenamefont {Xiong},
  \citenamefont {Lo}, \citenamefont {Zhang}, \citenamefont {Feng},\ and\
  \citenamefont {Nori}}]{Xiong15}%
  \BibitemOpen
  \bibfield  {author} {\bibinfo {author} {\bibfnamefont {H.-N.}\ \bibnamefont
  {Xiong}}, \bibinfo {author} {\bibfnamefont {P.-Y.}\ \bibnamefont {Lo}},
  \bibinfo {author} {\bibfnamefont {W.-M.}\ \bibnamefont {Zhang}}, \bibinfo
  {author} {\bibfnamefont {D.~H.}\ \bibnamefont {Feng}}, \ and\ \bibinfo
  {author} {\bibfnamefont {F.}~\bibnamefont {Nori}},\ }\href@noop {} {\bibfield
   {journal} {\bibinfo  {journal} {Scientific reports}\ }\textbf {\bibinfo
  {volume} {5}} (\bibinfo {year} {2015})}\BibitemShut {NoStop}%
\bibitem [{\citenamefont {Zhang}\ \emph
  {et~al.}(2012{\natexlab{b}})\citenamefont {Zhang}, \citenamefont {Lo},
  \citenamefont {Xiong}, \citenamefont {Tu},\ and\ \citenamefont
  {Nori}}]{Zhang12prl}%
  \BibitemOpen
  \bibfield  {author} {\bibinfo {author} {\bibfnamefont {W.-M.}\ \bibnamefont
  {Zhang}}, \bibinfo {author} {\bibfnamefont {P.-Y.}\ \bibnamefont {Lo}},
  \bibinfo {author} {\bibfnamefont {H.-N.}\ \bibnamefont {Xiong}}, \bibinfo
  {author} {\bibfnamefont {M.~W.-Y.}\ \bibnamefont {Tu}}, \ and\ \bibinfo
  {author} {\bibfnamefont {F.}~\bibnamefont {Nori}},\ }\href {\doibase
  10.1103/PhysRevLett.109.170402} {\bibfield  {journal} {\bibinfo  {journal}
  {Phys. Rev. Lett.}\ }\textbf {\bibinfo {volume} {109}},\ \bibinfo {pages}
  {170402} (\bibinfo {year} {2012}{\natexlab{b}})}\BibitemShut {NoStop}%
\bibitem [{\citenamefont {Jin}\ \emph {et~al.}(2010)\citenamefont {Jin},
  \citenamefont {Tu}, \citenamefont {Zhang},\ and\ \citenamefont
  {Yan}}]{Jin10}%
  \BibitemOpen
  \bibfield  {author} {\bibinfo {author} {\bibfnamefont {J.}~\bibnamefont
  {Jin}}, \bibinfo {author} {\bibfnamefont {M.~W.-Y.}\ \bibnamefont {Tu}},
  \bibinfo {author} {\bibfnamefont {W.-M.}\ \bibnamefont {Zhang}}, \ and\
  \bibinfo {author} {\bibfnamefont {Y.}~\bibnamefont {Yan}},\ }\href@noop {}
  {\bibfield  {journal} {\bibinfo  {journal} {New Journal of Physics}\ }\textbf
  {\bibinfo {volume} {12}},\ \bibinfo {pages} {083013} (\bibinfo {year}
  {2010})}\BibitemShut {NoStop}%
\bibitem [{\citenamefont {Bravyi}\ and\ \citenamefont
  {Kitaev}(2002)}]{Bravyi02}%
  \BibitemOpen
  \bibfield  {author} {\bibinfo {author} {\bibfnamefont {S.~B.}\ \bibnamefont
  {Bravyi}}\ and\ \bibinfo {author} {\bibfnamefont {A.~Y.}\ \bibnamefont
  {Kitaev}},\ }\href {\doibase http://dx.doi.org/10.1006/aphy.2002.6254}
  {\bibfield  {journal} {\bibinfo  {journal} {Annals of Physics}\ }\textbf
  {\bibinfo {volume} {298}},\ \bibinfo {pages} {210 } (\bibinfo {year}
  {2002})}\BibitemShut {NoStop}%
\bibitem [{\citenamefont {Haug}\ and\ \citenamefont {Jauho}(2008)}]{Jauhobook}%
  \BibitemOpen
  \bibfield  {author} {\bibinfo {author} {\bibfnamefont {H.}~\bibnamefont
  {Haug}}\ and\ \bibinfo {author} {\bibfnamefont {A.-P.}\ \bibnamefont
  {Jauho}},\ }\href@noop {} {\emph {\bibinfo {title} {{Quantum Kinetics in
  Transport and Optics of Semiconductors}}}},\ \bibinfo {edition} {2nd}\ ed.\
  (\bibinfo  {publisher} {Springer},\ \bibinfo {address} {Berlin Heidelberg},\
  \bibinfo {year} {2008})\BibitemShut {NoStop}%
\bibitem [{\citenamefont {Schaller}(2014)}]{Schallerbook}%
  \BibitemOpen
  \bibfield  {author} {\bibinfo {author} {\bibfnamefont {G.}~\bibnamefont
  {Schaller}},\ }\href@noop {} {\emph {\bibinfo {title} {{Open Quantum Systems
  Far From Equilibrium}}}},\ \bibinfo {edition} {1st}\ ed.\ (\bibinfo
  {publisher} {Springer},\ \bibinfo {address} {Berlin Heidelberg},\ \bibinfo
  {year} {2014})\BibitemShut {NoStop}%
\bibitem [{Note1()}]{Note1}%
  \BibitemOpen
  \bibinfo {note} {This tool has been applied in problems involving interacting
  quantum dynamics, one being the classic work of Haldane~\cite {Haldane80} who
  show the equivalence between the $XXZ$ model for a Heisenberg-Ising chain and
  interacting spinless fermions. Other example of the application of this
  transformation on the physics of strong correlated systems are the use of a
  generalized of the transformation in the solution of some cases of lattice
  models~\cite {Batista01}.}\BibitemShut {Stop}%
\bibitem [{Note2()}]{Note2}%
  \BibitemOpen
  \bibinfo {note} {Additionally, $\protect \mathrm {Tr}_n\protect \{ \rho _n
  \protect \}=\protect \mathrm {Tr}_n\protect \{ \sigma _z^{\otimes K_n} \rho
  _n \sigma _z^{\otimes K_n } \protect \}=1$}\BibitemShut {NoStop}%
\bibitem [{\citenamefont {Havel}(2003)}]{Havel03}%
  \BibitemOpen
  \bibfield  {author} {\bibinfo {author} {\bibfnamefont {T.~F.}\ \bibnamefont
  {Havel}},\ }\href {\doibase 10.1063/1.1518555} {\bibfield  {journal}
  {\bibinfo  {journal} {J. Math. Phys.}\ }\textbf {\bibinfo {volume} {44}},\
  \bibinfo {pages} {534} (\bibinfo {year} {2003})}\BibitemShut {NoStop}%
\bibitem [{\citenamefont {Xie}\ \emph {et~al.}(1995)\citenamefont {Xie},
  \citenamefont {Madhukar}, \citenamefont {Chen},\ and\ \citenamefont
  {Kobayashi}}]{Kobayashi95}%
  \BibitemOpen
  \bibfield  {author} {\bibinfo {author} {\bibfnamefont {Q.}~\bibnamefont
  {Xie}}, \bibinfo {author} {\bibfnamefont {A.}~\bibnamefont {Madhukar}},
  \bibinfo {author} {\bibfnamefont {P.}~\bibnamefont {Chen}}, \ and\ \bibinfo
  {author} {\bibfnamefont {N.~P.}\ \bibnamefont {Kobayashi}},\ }\href@noop {}
  {\bibfield  {journal} {\bibinfo  {journal} {Phys. Rev. Lett.}\ }\textbf
  {\bibinfo {volume} {75}},\ \bibinfo {pages} {2542} (\bibinfo {year}
  {1995})}\BibitemShut {NoStop}%
\bibitem [{\citenamefont {Tarucha}\ \emph {et~al.}(1996)\citenamefont
  {Tarucha}, \citenamefont {Austing}, \citenamefont {Honda}, \citenamefont
  {van~der Hage},\ and\ \citenamefont {Kouwenhoven}}]{Tarucha96}%
  \BibitemOpen
  \bibfield  {author} {\bibinfo {author} {\bibfnamefont {S.}~\bibnamefont
  {Tarucha}}, \bibinfo {author} {\bibfnamefont {D.~G.}\ \bibnamefont
  {Austing}}, \bibinfo {author} {\bibfnamefont {T.}~\bibnamefont {Honda}},
  \bibinfo {author} {\bibfnamefont {R.~J.}\ \bibnamefont {van~der Hage}}, \
  and\ \bibinfo {author} {\bibfnamefont {L.~P.}\ \bibnamefont {Kouwenhoven}},\
  }\href@noop {} {\bibfield  {journal} {\bibinfo  {journal} {Phys. Rev. Lett.}\
  }\textbf {\bibinfo {volume} {77}},\ \bibinfo {pages} {3613} (\bibinfo {year}
  {1996})}\BibitemShut {NoStop}%
\bibitem [{\citenamefont {Fujisawa}\ \emph {et~al.}(1998)\citenamefont
  {Fujisawa}, \citenamefont {Oosterkamp}, \citenamefont {van~der Wiel},
  \citenamefont {Broer}, \citenamefont {Aguado}, \citenamefont {Tarucha},\ and\
  \citenamefont {Kouwenhoven}}]{Fujisawa98}%
  \BibitemOpen
  \bibfield  {author} {\bibinfo {author} {\bibfnamefont {T.}~\bibnamefont
  {Fujisawa}}, \bibinfo {author} {\bibfnamefont {T.~H.}\ \bibnamefont
  {Oosterkamp}}, \bibinfo {author} {\bibfnamefont {W.~G.}\ \bibnamefont
  {van~der Wiel}}, \bibinfo {author} {\bibfnamefont {B.~W.}\ \bibnamefont
  {Broer}}, \bibinfo {author} {\bibfnamefont {R.}~\bibnamefont {Aguado}},
  \bibinfo {author} {\bibfnamefont {S.}~\bibnamefont {Tarucha}}, \ and\
  \bibinfo {author} {\bibfnamefont {L.~P.}\ \bibnamefont {Kouwenhoven}},\
  }\href@noop {} {\bibfield  {journal} {\bibinfo  {journal} {Science}\ }\textbf
  {\bibinfo {volume} {282}},\ \bibinfo {pages} {932} (\bibinfo {year}
  {1998})}\BibitemShut {NoStop}%
\bibitem [{\citenamefont {Oosterkamp}\ \emph {et~al.}(1998)\citenamefont
  {Oosterkamp}, \citenamefont {Fujisawa}, \citenamefont {van~der Wiel},
  \citenamefont {Ishibashi}, \citenamefont {Hijman}, \citenamefont {Tarucha},\
  and\ \citenamefont {Kouwenhoven}}]{Oosterkamp98}%
  \BibitemOpen
  \bibfield  {author} {\bibinfo {author} {\bibfnamefont {T.~H.}\ \bibnamefont
  {Oosterkamp}}, \bibinfo {author} {\bibfnamefont {T.}~\bibnamefont
  {Fujisawa}}, \bibinfo {author} {\bibfnamefont {W.~G.}\ \bibnamefont {van~der
  Wiel}}, \bibinfo {author} {\bibfnamefont {K.}~\bibnamefont {Ishibashi}},
  \bibinfo {author} {\bibfnamefont {R.~V.}\ \bibnamefont {Hijman}}, \bibinfo
  {author} {\bibfnamefont {S.}~\bibnamefont {Tarucha}}, \ and\ \bibinfo
  {author} {\bibfnamefont {L.~P.}\ \bibnamefont {Kouwenhoven}},\ }\href@noop {}
  {\bibfield  {journal} {\bibinfo  {journal} {Science}\ }\textbf {\bibinfo
  {volume} {395}},\ \bibinfo {pages} {873} (\bibinfo {year}
  {1998})}\BibitemShut {NoStop}%
\bibitem [{\citenamefont {Hayashi}\ \emph {et~al.}(2003)\citenamefont
  {Hayashi}, \citenamefont {Fujisawa}, \citenamefont {Cheong}, \citenamefont
  {Jeong},\ and\ \citenamefont {Hirayama}}]{Hayashi03}%
  \BibitemOpen
  \bibfield  {author} {\bibinfo {author} {\bibfnamefont {T.}~\bibnamefont
  {Hayashi}}, \bibinfo {author} {\bibfnamefont {T.}~\bibnamefont {Fujisawa}},
  \bibinfo {author} {\bibfnamefont {H.~D.}\ \bibnamefont {Cheong}}, \bibinfo
  {author} {\bibfnamefont {Y.~H.}\ \bibnamefont {Jeong}}, \ and\ \bibinfo
  {author} {\bibfnamefont {Y.}~\bibnamefont {Hirayama}},\ }\href@noop {}
  {\bibfield  {journal} {\bibinfo  {journal} {Phys. Rev. Lett.}\ }\textbf
  {\bibinfo {volume} {91}},\ \bibinfo {pages} {226804} (\bibinfo {year}
  {2003})}\BibitemShut {NoStop}%
\bibitem [{\citenamefont {Shinkai}\ \emph {et~al.}(2009)\citenamefont
  {Shinkai}, \citenamefont {Hayashi}, \citenamefont {Ota},\ and\ \citenamefont
  {Fujisawa}}]{Shinkai09}%
  \BibitemOpen
  \bibfield  {author} {\bibinfo {author} {\bibfnamefont {G.}~\bibnamefont
  {Shinkai}}, \bibinfo {author} {\bibfnamefont {T.}~\bibnamefont {Hayashi}},
  \bibinfo {author} {\bibfnamefont {T.}~\bibnamefont {Ota}}, \ and\ \bibinfo
  {author} {\bibfnamefont {T.}~\bibnamefont {Fujisawa}},\ }\href@noop {}
  {\bibfield  {journal} {\bibinfo  {journal} {Phys. Rev. Lett.}\ }\textbf
  {\bibinfo {volume} {103}},\ \bibinfo {pages} {056802} (\bibinfo {year}
  {2009})}\BibitemShut {NoStop}%
\bibitem [{\citenamefont {Shinkai}\ \emph {et~al.}(2007)\citenamefont
  {Shinkai}, \citenamefont {Hayashi}, \citenamefont {Hirayama},\ and\
  \citenamefont {Fujisawa}}]{Shinkai07}%
  \BibitemOpen
  \bibfield  {author} {\bibinfo {author} {\bibfnamefont {G.}~\bibnamefont
  {Shinkai}}, \bibinfo {author} {\bibfnamefont {T.}~\bibnamefont {Hayashi}},
  \bibinfo {author} {\bibfnamefont {Y.}~\bibnamefont {Hirayama}}, \ and\
  \bibinfo {author} {\bibfnamefont {T.}~\bibnamefont {Fujisawa}},\ }\href@noop
  {} {\bibfield  {journal} {\bibinfo  {journal} {Appl. Phys. Lett.}\ }\textbf
  {\bibinfo {volume} {90}},\ \bibinfo {pages} {103116} (\bibinfo {year}
  {2007})}\BibitemShut {NoStop}%
\bibitem [{\citenamefont {Fujisawa}\ \emph {et~al.}(2011)\citenamefont
  {Fujisawa}, \citenamefont {Shinkai}, \citenamefont {Hayashi},\ and\
  \citenamefont {Ota}}]{Fujisawa11}%
  \BibitemOpen
  \bibfield  {author} {\bibinfo {author} {\bibfnamefont {T.}~\bibnamefont
  {Fujisawa}}, \bibinfo {author} {\bibfnamefont {G.}~\bibnamefont {Shinkai}},
  \bibinfo {author} {\bibfnamefont {T.}~\bibnamefont {Hayashi}}, \ and\
  \bibinfo {author} {\bibfnamefont {T.}~\bibnamefont {Ota}},\ }\href@noop {}
  {\bibfield  {journal} {\bibinfo  {journal} {Phys. E}\ }\textbf {\bibinfo
  {volume} {43}},\ \bibinfo {pages} {730734} (\bibinfo {year}
  {2011})}\BibitemShut {NoStop}%
\bibitem [{\citenamefont {Fanchini}\ \emph {et~al.}(2010)\citenamefont
  {Fanchini}, \citenamefont {Castelano},\ and\ \citenamefont
  {Caldeira}}]{Fanchini10}%
  \BibitemOpen
  \bibfield  {author} {\bibinfo {author} {\bibfnamefont {F.}~\bibnamefont
  {Fanchini}}, \bibinfo {author} {\bibfnamefont {L.~K.}\ \bibnamefont
  {Castelano}}, \ and\ \bibinfo {author} {\bibfnamefont {A.~O.}\ \bibnamefont
  {Caldeira}},\ }\href@noop {} {\bibfield  {journal} {\bibinfo  {journal} {New
  J. Phys.}\ }\textbf {\bibinfo {volume} {12}},\ \bibinfo {pages} {073009}
  (\bibinfo {year} {2010})}\BibitemShut {NoStop}%
\bibitem [{\citenamefont {Oliveira}\ and\ \citenamefont
  {Sanz}(2015)}]{Oliveira15}%
  \BibitemOpen
  \bibfield  {author} {\bibinfo {author} {\bibfnamefont {P.}~\bibnamefont
  {Oliveira}}\ and\ \bibinfo {author} {\bibfnamefont {L.}~\bibnamefont
  {Sanz}},\ }\href@noop {} {\bibfield  {journal} {\bibinfo  {journal} {Ann.
  Physics}\ }\textbf {\bibinfo {volume} {356}},\ \bibinfo {pages} {244}
  (\bibinfo {year} {2015})}\BibitemShut {NoStop}%
\bibitem [{\citenamefont {Wootters}(1998)}]{Wootters98}%
  \BibitemOpen
  \bibfield  {author} {\bibinfo {author} {\bibfnamefont {W.~K.}\ \bibnamefont
  {Wootters}},\ }\href@noop {} {\bibfield  {journal} {\bibinfo  {journal}
  {Phys. Rev. Lett.}\ }\textbf {\bibinfo {volume} {80}},\ \bibinfo {pages}
  {2245} (\bibinfo {year} {1998})}\BibitemShut {NoStop}%
\bibitem [{\citenamefont {Hill}\ and\ \citenamefont {Wootters}(1997)}]{Hill97}%
  \BibitemOpen
  \bibfield  {author} {\bibinfo {author} {\bibfnamefont {S.}~\bibnamefont
  {Hill}}\ and\ \bibinfo {author} {\bibfnamefont {W.~K.}\ \bibnamefont
  {Wootters}},\ }\href {\doibase 10.1103/PhysRevLett.78.5022} {\bibfield
  {journal} {\bibinfo  {journal} {Phys. Rev. Lett.}\ }\textbf {\bibinfo
  {volume} {78}},\ \bibinfo {pages} {5022} (\bibinfo {year}
  {1997})}\BibitemShut {NoStop}%
\bibitem [{\citenamefont {Yu}\ and\ \citenamefont
  {Eberly}(2006{\natexlab{a}})}]{Yu06a}%
  \BibitemOpen
  \bibfield  {author} {\bibinfo {author} {\bibfnamefont {T.}~\bibnamefont
  {Yu}}\ and\ \bibinfo {author} {\bibfnamefont {J.~H.}\ \bibnamefont
  {Eberly}},\ }\href {\doibase 10.1103/PhysRevLett.97.140403} {\bibfield
  {journal} {\bibinfo  {journal} {Phys. Rev. Lett.}\ }\textbf {\bibinfo
  {volume} {97}},\ \bibinfo {pages} {140403} (\bibinfo {year}
  {2006}{\natexlab{a}})}\BibitemShut {NoStop}%
\bibitem [{\citenamefont {Yu}\ and\ \citenamefont
  {Eberly}(2006{\natexlab{b}})}]{Yu06b}%
  \BibitemOpen
  \bibfield  {author} {\bibinfo {author} {\bibfnamefont {T.}~\bibnamefont
  {Yu}}\ and\ \bibinfo {author} {\bibfnamefont {J.}~\bibnamefont {Eberly}},\
  }\href {\doibase http://dx.doi.org/10.1016/j.optcom.2006.01.061} {\bibfield
  {journal} {\bibinfo  {journal} {Optics Communications}\ }\textbf {\bibinfo
  {volume} {264}},\ \bibinfo {pages} {393 } (\bibinfo {year}
  {2006}{\natexlab{b}})},\ \bibinfo {note} {quantum Control of Light and
  Matter}\BibitemShut {NoStop}%
\bibitem [{\citenamefont {Almeida}\ \emph {et~al.}(2007)\citenamefont
  {Almeida}, \citenamefont {de~Melo}, \citenamefont {Hor-Meyll}, \citenamefont
  {Salles}, \citenamefont {Walborn}, \citenamefont {Souto~Ribeiro},\ and\
  \citenamefont {Davidovich}}]{Almeida07}%
  \BibitemOpen
  \bibfield  {author} {\bibinfo {author} {\bibfnamefont {M.}~\bibnamefont
  {Almeida}}, \bibinfo {author} {\bibfnamefont {F.}~\bibnamefont {de~Melo}},
  \bibinfo {author} {\bibfnamefont {M.}~\bibnamefont {Hor-Meyll}}, \bibinfo
  {author} {\bibfnamefont {A.}~\bibnamefont {Salles}}, \bibinfo {author}
  {\bibfnamefont {S.}~\bibnamefont {Walborn}}, \bibinfo {author} {\bibfnamefont
  {P.}~\bibnamefont {Souto~Ribeiro}}, \ and\ \bibinfo {author} {\bibfnamefont
  {L.}~\bibnamefont {Davidovich}},\ }\href@noop {} {\bibfield  {journal}
  {\bibinfo  {journal} {Science}\ }\textbf {\bibinfo {volume} {316}},\ \bibinfo
  {pages} {579} (\bibinfo {year} {2007})}\BibitemShut {NoStop}%
\bibitem [{\citenamefont {Haldane}(1980)}]{Haldane80}%
  \BibitemOpen
  \bibfield  {author} {\bibinfo {author} {\bibfnamefont {F.~D.~M.}\
  \bibnamefont {Haldane}},\ }\href {\doibase 10.1103/PhysRevLett.45.1358}
  {\bibfield  {journal} {\bibinfo  {journal} {Phys. Rev. Lett.}\ }\textbf
  {\bibinfo {volume} {45}},\ \bibinfo {pages} {1358} (\bibinfo {year}
  {1980})}\BibitemShut {NoStop}%
\bibitem [{\citenamefont {Batista}\ and\ \citenamefont
  {Ortiz}(2001)}]{Batista01}%
  \BibitemOpen
  \bibfield  {author} {\bibinfo {author} {\bibfnamefont {C.~D.}\ \bibnamefont
  {Batista}}\ and\ \bibinfo {author} {\bibfnamefont {G.}~\bibnamefont
  {Ortiz}},\ }\href {\doibase 10.1103/PhysRevLett.86.1082} {\bibfield
  {journal} {\bibinfo  {journal} {Phys. Rev. Lett.}\ }\textbf {\bibinfo
  {volume} {86}},\ \bibinfo {pages} {1082} (\bibinfo {year}
  {2001})}\BibitemShut {NoStop}%
\end{thebibliography}
\end{document}